\documentclass{vldb}

\usepackage{graphicx}
\usepackage{balance}

\usepackage{booktabs} 
\usepackage{listings}
\usepackage{amsmath}
\usepackage{algorithm}
\usepackage[noend]{algpseudocode}
\usepackage[lofdepth,lotdepth]{subfig}
\usepackage{enumitem}

\usepackage{url}
\newtheorem{definition}{Definition}

\newcommand{\eat}[1]{}
\newcommand{\mytilde}{\raise.17ex\hbox{$\scriptstyle\mathtt{\sim}$}}
\newlength{\oldtextfloatsep}

\vldbTitle{Automating Distributed Tiered Storage Management in Cluster Computing}
\vldbAuthors{Herodotos Herodotou and Elena Kakoulli}
\vldbDOI{https://doi.org/10.14778/3357377.3357381}
\vldbVolume{13}
\vldbNumber{1}
\vldbYear{2019}

\begin{document}


\title{Automating Distributed Tiered Storage Management in Cluster Computing}



%

\numberofauthors{2}

\author{
%
%
\alignauthor
Herodotos Herodotou\\
       \affaddr{Cyprus University of Technology}\\
       \affaddr{Limassol, Cyprus}\\
       \email{herodotos.herodotou@cut.ac.cy}
\alignauthor
Elena Kakoulli\\
       \affaddr{Cyprus University of Technology}\\
       \affaddr{Limassol, Cyprus}\\
       \email{elena.kakoulli@cut.ac.cy}
}

\date{1 April 2019}

\maketitle

\begin{abstract}

Data-intensive platforms such as Hadoop and Spark are routinely used to process massive amounts of data residing on distributed file systems like HDFS.
Increasing memory sizes and new hardware technologies (e.g., NVRAM, SSDs) have recently led to the introduction of storage tiering in such settings.
However, users are now burdened with the additional complexity of managing the multiple storage tiers 
and the data residing on them while trying to optimize their workloads.
In this paper, we develop a general framework for automatically moving data across the available storage tiers in distributed file systems.
Moreover, we employ machine learning for tracking and predicting file access patterns, which we use to decide when and which data to move up or down the storage tiers for increasing system performance.
Our approach uses incremental learning to dynamically refine the models with new file accesses, allowing them to naturally adjust and adapt to workload changes over time.
Our extensive evaluation using realistic workloads derived from Facebook and CMU traces compares our approach with several other policies and showcases significant benefits in terms of both workload performance and cluster efficiency.

\end{abstract}

\section{Introduction}
\label{sec:intro}

Data-intensive analytic applications for business intelligence, social network analysis, and scientific data processing are routinely executed on Big Data platforms such as Hadoop YARN \cite{yarn-socc13} and Spark \cite{spark-rds-nsdi12}, while processing massive amounts of data residing in distributed file systems such as HDFS \cite{hdfs-msst10}.
Such applications tend to spend significant fractions of their overall execution in performing I/O \cite{supercomputerIO_hpdc17}.
Larger memory sizes as well as new hardware technologies such as Non-Volatile RAM (NVRAM) and Solid-State Drives (SSDs) are now commonly utilized for increasing I/O performance.
In particular, in-memory distributed file systems like Alluxio \cite{alluxio-webpage} and GridGain \cite{gridgain-webpage} are used for storing or caching HDFS data in memory.
HDFS has also added support for caching input files internally \cite{hdfs-caching-webpage}.
NVRAM and SSDs have been used as the storage layer for distributed systems \cite{skimpystash-sigmod11, ssd-hadoop-cf13, one-pass-mapreduce-sigmod11} as well as in shared storage systems (e.g., ReCA \cite{reca-tpds18}, Hermes \cite{hermes-hpdc18}).
Recently, the OctopusFS distributed file system \cite{octopusfs-sigmod17} introduced fine-grained tiering in compute clusters via storing file replicas on the various storage media (e.g., memory, SSDs, HDDs) that are locally attached on the cluster nodes.
At the same time, HDFS generalized its architecture to support storage media other than HDDs, including memory and SSDs \cite{hdfs-jira-hetestor}.

As multiple storage tiers are added into distributed file systems (and storage systems in general), \textit{the complexity of data movement across the tiers increases significantly}, making it harder to take advantage of the higher I/O performance offered by the system \cite{hermes-hpdc18}.
Most aforementioned systems expose APIs for data caching or movement to application developers and data analysts.
For example, an HDFS application can issue requests to cache files
and directories. 
However, when the cache gets full, no additional caching requests will be served until the application \textit{manually} uncaches some files \cite{hdfs-caching-webpage}.
Similarly, OctopusFS offers a placement policy for determining how to initially store the data across the storage tiers but lacks any features for automatically moving data afterwards.
Other systems like Alluxio \cite{alluxio-webpage} and GridGain \cite{gridgain-webpage} implement basic policies for removing data from memory when full, such as LRU (Least Recently Used) \cite{web-cache-survey-ijasca11}.
However, such policies are known to under-perform in the big data setting as they were initially designed for evicting fixed-size pages from buffer caches \cite{pacman-nsdi12,web-cache-survey-ijasca11}.
Furthermore, these systems do not offer any cache admission policies; i.e., they will place all data in the cache upon access, without any regards for the current state of the system, the data size, or any workload patterns.

Overall, the lack of automated data movement across storage tiers places a significant burden on users or system administrators, who are tasked with optimizing various types of data analytics workloads \cite{hermes-hpdc18, self-star-icac04}.
The analysis of production workloads from Facebook, Cloudera, and MS Bing \cite{pacman-nsdi12, mr-workloads-vldb12} has shown that many jobs (small and large alike) exhibit data re-access patterns both in the short-term (within hours) and the long-term (daily, weekly, or monthly).
Thus, identifying the reused data and keeping them in higher tiers can yield significant performance benefits \cite{pacman-nsdi12}.
In addition, data access patterns can change over time 
with the addition and removal of users and jobs,
leading to an increased need for adapting to these changes accurately and efficiently \cite{self-star-icac04}.
Hence, it is imperative for tiered storage systems to include automated data management capabilities for improving cluster efficiency and application performance.

In this paper, we propose a \textit{general framework for automated tiered storage management in distributed file systems}.
Specifically, our framework can be used for orchestrating data management policies for adaptively deciding (i) when and which data to retain or move to higher storage tiers for improved read performance and (ii) when and which data to move to lower tiers for freeing scarce resources.
We show its generalization by implementing several conventional cache eviction and admission policies \cite{web-cache-survey-ijasca11}, related policies from recent literature \cite{pacman-nsdi12, big-sql-caching-socc16}, as well as our own policies.

Furthermore, we propose the use of \textit{machine learning (ML) for tracking and predicting file access patterns} in the system.
In particular, we employ light-weight \textit{gradient boosted trees} \cite{xgboost-kdd16} to learn how files are accessed by the current workload and use the generated models to drive our automated file system policies.
Our approach uses \textit{incremental learning} to dynamically refine the models with new file accesses as they become available, allowing the models to naturally adjust and adapt to workload changes over time.

Our work lies at the intersection of distributed storage systems, hierarchical storage management (HSM) solutions, and caching, with a humbling amount of related work.
Yet, to the best of our knowledge, \textit{we propose and implement a new, fully automated, and adaptive approach to tiered storage management in distributed file and storage systems}.
Our ML-based approach is also different from previous approaches in HSM and caching (a detailed comparison is provided in Section \ref{sec:prelim}).
Finally, we have implemented our approach in an existing distributed file system, namely OctopusFS \cite{octopusfs-sigmod17}, which is a backwards-compatible extension of HDFS \cite{hdfs-msst10}.

\vspace{1pt}
In summary, the key contributions of this paper are:
\begin{enumerate}[leftmargin=*, itemsep=0pt]
\item The design and implementation of a general framework for automatically managing storage tiers in distributed file systems.
\item An online, adaptive machine learning-based policy for predicting file access patterns and dynamically moving data among storage tiers.
\item An extensive evaluation using realistic workloads derived from Facebook and CMU traces, showcasing significant benefits for workload performance and cluster efficiency.
\end{enumerate}

\noindent
The paper is organized as follows. 
Section \ref{sec:prelim} presents an overview and comparison with existing related work.
Section \ref{sec:overview} discusses the proposed tiered storage management framework and Section \ref{sec:mlapproach} formulates the ML models used for predicting file access patterns.
Sections \ref{sec:downgrade} and \ref{sec:upgrade} outline several policies for moving files down and up the storage hierarchy, respectively. 
The experimental evaluation is presented in Section \ref{sec:evaluation}, while Section \ref{sec:conclusions} concludes the paper.

\section{Background and Related Work}
\label{sec:prelim}

In this section, we provide a brief background of storage tiering and compare previous research with our work.

\subsection{Distributed File Systems and Tiering}
\label{sec:prelim:dfs}

\begin{figure}
	\centering
	\includegraphics[width=0.47\textwidth]{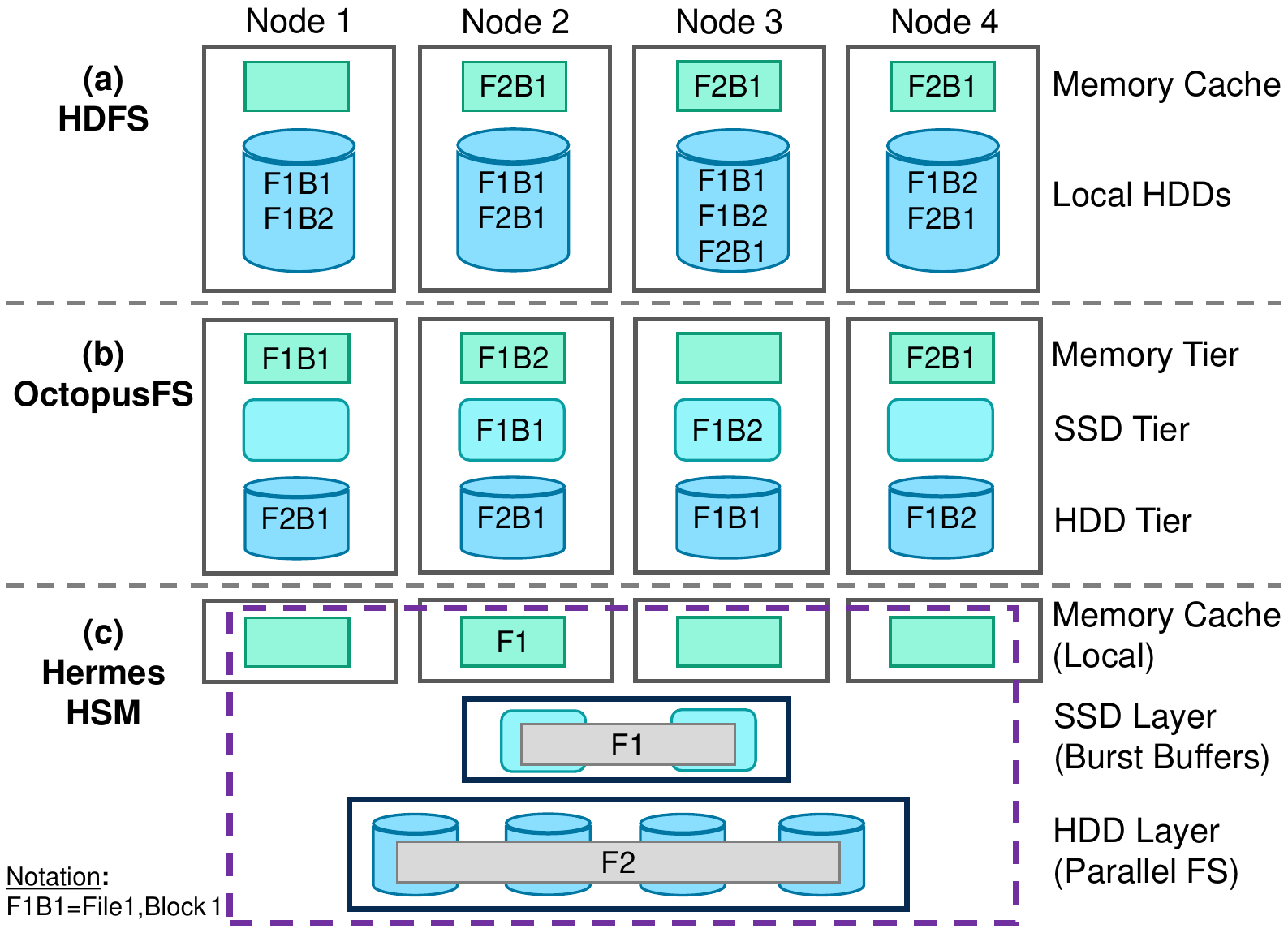}
	\vspace{1.5ex}
	\caption{Data placement for three tiered storage systems}
	\label{fig:data-placement}
	\vspace{-1ex}
\end{figure}

Distributed file systems like \textit{HDFS} \cite{hdfs-msst10} serve the current generation of Big Data platforms \cite{yarn-socc13, spark-rds-nsdi12} via storing files on locally-attached hard disk drives (HDDs) on the cluster nodes.
The files are typically broken down into large blocks (e.g., 128MB), which are then replicated and distributed in the cluster, as shown in Figure \ref{fig:data-placement}(a).
HDFS has recently taken significant steps toward tiered storage via (i) enabling the HDFS cache \cite{hdfs-caching-webpage}, and (ii) supporting heterogeneous storage types such as SSDs and memory \cite{hdfs-jira-hetestor}.
In the former, there is no support for automatically caching or uncaching files, while in the latter, there is support for a limited number of static policies for storing files on specific tiers \cite{hdfs-jira-hetestor}.
A caching middleware has been proposed for using local SSDs as a read-only cache of local HDDs in HDFS \cite{caching-middleware-ijbgi16},
which uses a heuristic file-placement algorithm to improve the cache but assumes that a queue of requested files is known in advanced.

\textit{hatS} \cite{hats-ccgrid14} and \textit{OctopusFS} \cite{octopusfs-sigmod17} extended HDFS to support fine-grained storage tiering based on which file blocks are replicated and stored across both the cluster nodes and the storage tiers (see Figure \ref{fig:data-placement}(b)).
hatS proposed a simple rule-based data placement policy, whereas OctopusFS developed one based on a multi-objective problem formulation for deciding how the file blocks should be distributed in the cluster in order to maximize performance, fault tolerance, and data and load balancing.
However, neither hatS nor OctopusFS support any policies for automatically moving file replicas between the storage tiers.
To exploit larger memories, in-memory file systems such as \textit{Alluxio} \cite{alluxio-webpage} and \textit{GridGain} \cite{gridgain-webpage} can be used for storing or caching data in clusters.
In the Spark ecosystem, \textit{Resilient Distributed Datasets (RDDs)} are a distributed memory abstraction that allows applications to persist a specified dataset in memory for reuse, and use lineage for fault tolerance \cite{spark-rds-nsdi12}. 
While conventional cache eviction policies such as LRU are used in these systems, they rely on the user to manually cache the data.

\textit{PACMan} \cite{pacman-nsdi12} is a memory caching system that explores memory locality of data-intensive jobs.
PACMan implements two eviction policies: (i) \textit{LIFE} that minimizes job completion time by prioritizing small inputs, and (ii) \textit{LFU-F} that maximizes cluster efficiency by evicting less frequently accessed inputs. However, PACMan does not allow applications to specify hot data in memory for subsequent efficient accesses and does not implement cache admission policies.
\textit{Big SQL} \cite{big-sql-caching-socc16} is an SQL-on-hadoop system that utilizes HDFS cache for caching table partitions. 
Big SQL presents two algorithms, namely \textit{SLRU-K} and \textit{EXD}, that explore the tradeoff of caching objects based on recency and frequency of data accesses.
The two algorithms drive both cache eviction and admission policies but, unlike our approach, do not learn from file access patterns as the workload changes.

\subsection{Hierarchical Storage Management}
\label{sec:prelim:hsm}

{\em Hierarchical storage management (HSM)} solutions store data across storage layers that typically consists of arrays of tapes, compressed disks, and high-performance disks, while more recently memory and NVRAM are also exploited \cite{data-staging-ipdps15, hermes-hpdc18} (see Figure \ref{fig:data-placement}(c)).
HSM supports both \textit{tiering} (i.e., a file will only reside on one of the storage layers) and \textit{caching} (i.e., a copy of a file will be moved to the cache), but,
unlike distributed file systems, HSM does not offer any locality or storage-media awareness to higher-level applications.

The process of moving or copying files from one storage layer to another is typically based on predefined policies and parameters (e.g., low/high thresholds for disks capacities) \cite{data-staging-ipdps15, hermes-hpdc18}.
Other systems, such as \textit{DataWarp}, provides a script-based API for users to move the data between the SSD and the HDD layers \cite{datawarp-manual16}.
The \textit{ReCA} storage system uses the SSD layer as a dynamic cache \cite{reca-tpds18}.
Based on the incoming application I/Os, ReCA categorizes the workload in one of five types, and reconfigures the cache when detecting a change in the workload.
The \textit{DataSpaces} framework \cite{data-staging-ipdps15} exploits memory and SSDs to support dynamic data staging in HSM,
driven by user-provided hints about expected data read patterns.
\textit{Hermes} adds memory and NVRAM in the storage hierarchy and proposes three policies that can be manually configured by the user.
The cost models and tiering mechanisms used in prior approaches in HSM cannot be directly applied to analytics applications since they are designed to handle block level I/Os (e.g., 4--32 KB) for POSIX-style workloads (e.g., server, database, file systems) \cite{cast-hpdc15}.
In addition, our main approach for automatically moving data across the storage hierarchy is not based on parameter-driven or user-defined policies but rather on machine learning.

\begin{figure*}[t!]
	\centering
	\includegraphics[width=0.47\textwidth]{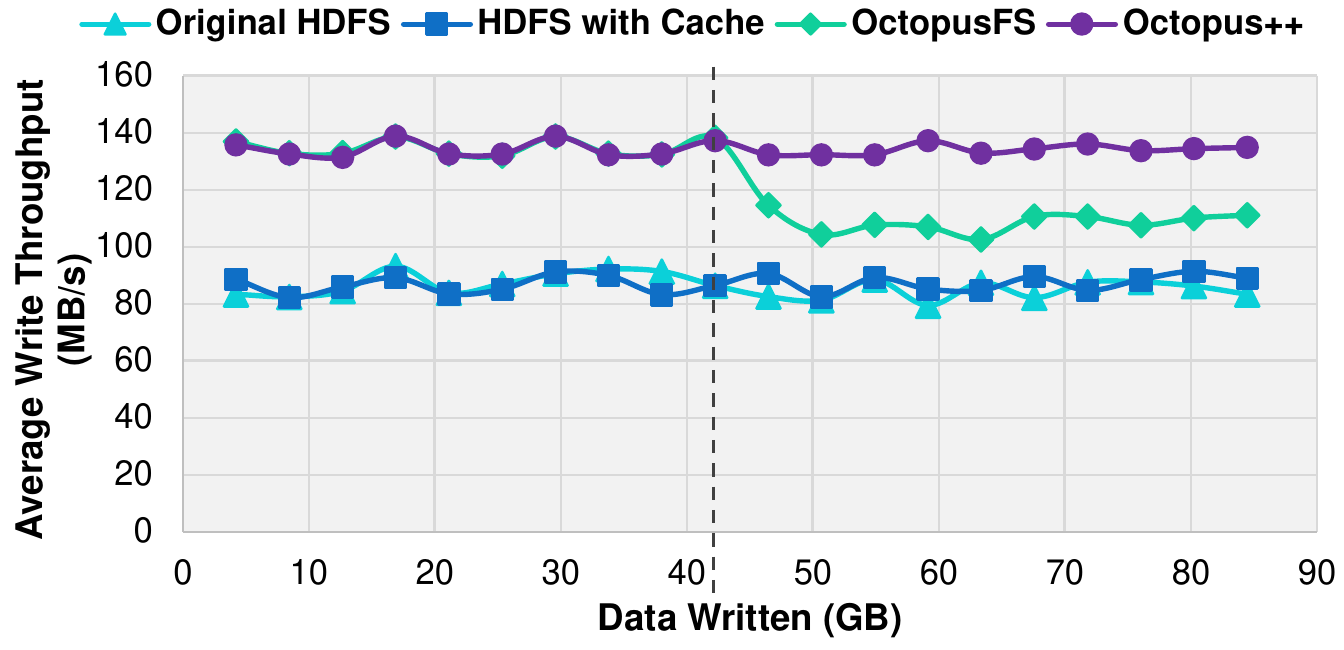}
	\hspace{6mm}
	\includegraphics[width=0.47\textwidth]{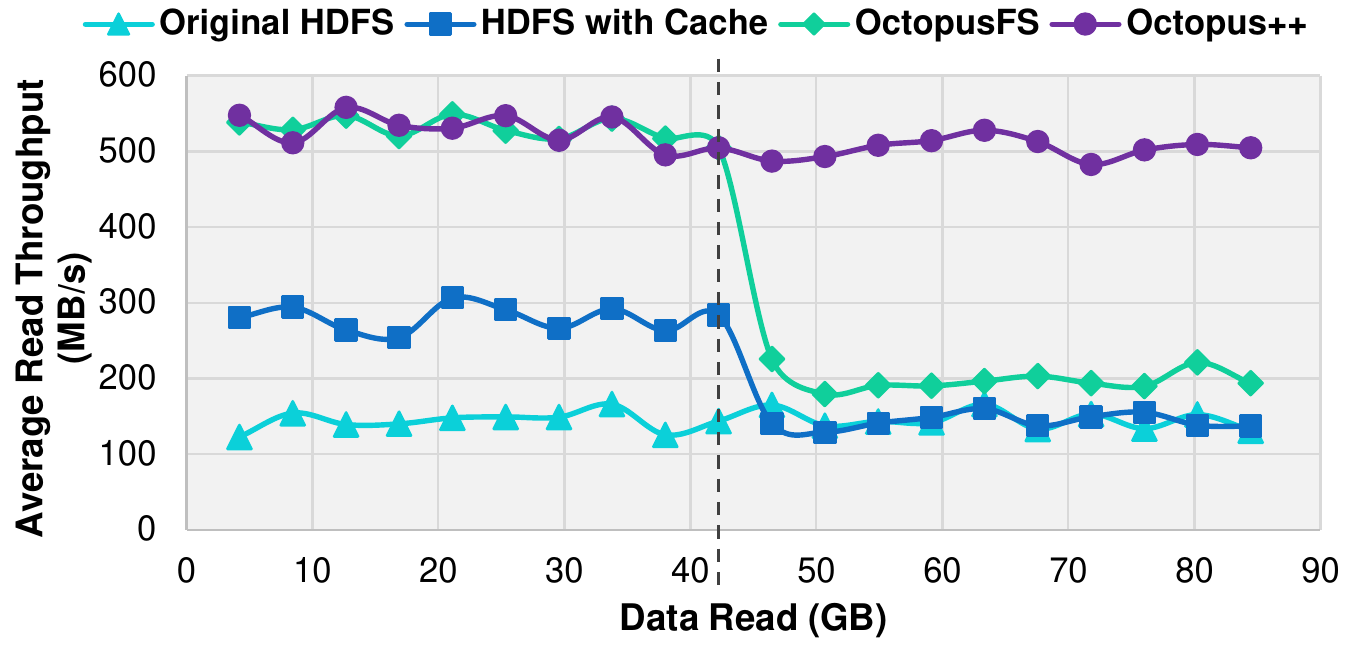}
	\vspace{0.5ex}
	\caption{(a) Average write and (b) read throughput per node for HDFS, HDFS with Cache, OctopusFS, and Octopus++}
	\label{fig:dfsio-thru}
	\vspace{-1.5ex}
\end{figure*}

\subsection{Caching}
\label{sec:prelim:cache}

Caching is a well-studied problem that appears in various contexts and discussed extensively in several surveys \cite{survey-cache-repl-csur03, web-cache-survey-ijasca11}.
We offer a quick overview of the area and highlight the most closely related work.
For CPU caches, virtual memory systems, and database buffer caches, there is extensive work on \textit{cache eviction} policies (e.g., LRU, LFU, ARC, MQ), which are classified as a) recency-based, b) frequency-based, c) size-based, d) function-based, and e) randomized \cite{survey-cache-repl-csur03}.
Other policies found in main memory databases attempt to identify hot and cold data, also based on access frequencies \cite{hot-cold-main-mem-db-icde13, cold-data-hyperdb-pvldb12, siberia-cold-data-pvldb14}.
Another recent policy augmented the Marker caching algorithm \cite{marker-algo-jalgo91} with a machine learned oracle for improving its competitive ratio \cite{augment-caching-ml-icml18}.
Unlike our approach, these policies operate on fixed-size pages and assume that every accessed page will be inserted into the cache.

Many caching policies have been developed for web caches that operate on variable size objects, including SIZE, Hyper-G, Greedy-Dual-Size, and Hybrid \cite{web-cache-survey-ijasca11}.
Most of these policies have been designed for improving the hit ratio in web accesses but that does not necessarily improve the performance in storage systems \cite{reca-tpds18}.
\textit{Machine learning} techniques such as logistic regression, artificial neural networks, genetic algorithms, random forests, and others have also been used for developing more intelligent web caching policies \cite{web-cache-survey-ijasca11, survey-nn-caching-iete09, edge-caching-survey-wireless18, caching-supervised-ml-ijasca14, web-caching-ml-nnw11}.
However, most of these approaches try to identify and predict relationships between web objects; for example, a visit to web page $X$ is typically followed by a visit to web page $Y$.
Other approaches try to capture associations between file attributes (e.g., owner, creation time, and permissions) and properties (e.g., access pattern, lifespan, and size) \cite{self-star-icac04}.
Another recent approach used neural networks to analyze the inter-relationships among web requests for making caching decisions \cite{snn-cache-ciss18}.
However, such relationships and associations are not expected to be present in big data analytics workloads and, hence, are not applicable in our setting.
Reinforcement learning has also been attempted by using multiple experts to select the best cache eviction policy to use at any given time, but these approaches are known to outperform only the static policies (e.g., LRU, LFU) and are computationally and memory expensive \cite{acme-caching-wdas02, adaptive-caching-anips03, lecar-hotstorage18}.

\textit{Web prefetching} predicts the web objects expected to be requested in the near future and brings them into the cache before users actually request them.
Existing prefetching approaches can be classified in two categories:
(i) content-based that rely on the analysis of web page contents, and 
(ii) history-based that rely on the observed page access patterns \cite{web-cache-survey-ijasca11}.
History-based approaches using Hidden Markov Models \cite{madhyastha1997input}, probability graphs \cite{griffioen1995performance}, and tries \cite{kroeger1999case} exploit either relationships between files or association rules from access patterns, similar to the intelligent web caching policies discussed above.
On the contrary, our approach builds a ML model based on recency, frequency, and size of file accesses in order to predict which files will be accessed soon or not accessed for a while (see Section \ref{sec:mlapproach}).

\subsection{Tiering in Cloud Storage Systems}
\label{sec:prelim:cloud}

The increased variety in cloud storage offerings such as network-attached block storage (e.g., Amazon EBS, Google Persistent Disk) and remote object/blob stores (e.g., Amazon S3, Google CS) has recently sparked interest in tiered storage for the Cloud \cite{cast-hpdc15, azure-adls-sigmod17}.
\textit{Frugal Cloud File System} \cite{frugal-fs-eurosys12} is a cloud-based file system that spans multiple cloud storage services and aims at minimizing operational cost via increasing or decreasing the amount of storage at different levels of the hierarchy.
In our setting, the amount of storage available is fixed and based on the storage media present in the cluster.
\textit{Azure Data Lake Store} (ADLS) \cite{azure-adls-sigmod17} is designed to support multiple storage tiers while data movements are managed by the system based on high-level policies.
ADLS currently offers two tiers: Cosmos Storage (locally-attached SSDs and HDDs) and Azure Blob Storage.
\textit{CAST} \cite{cast-hpdc15} offers a coarse-grained tiering approach in the Google Cloud for improving the performance of analytics workloads.
CAST relies on extensive offline workload profiling to construct job performance prediction models on different cloud storage services, and uses them to generate a storage provisioning and data placement plan.
In contrast, our approach does not require any offline training and can adapt to changing analytics workloads.

\section{Tiered Storage Management}
\label{sec:overview}

The main operations of distributed file systems, such as HDFS \cite{hdfs-msst10} and OctopusFS \cite{octopusfs-sigmod17}, are to store and retrieve \textit{files}, which are broken into large \textit{blocks}.
In HDFS, blocks are replicated 3 times by default and distributed across cluster nodes.
Replication offers three main benefits:
(1) it prevents data loss due to disk or node failures;
(2) it enables higher I/O rates since the same block can be read in parallel (from different replicas); and
(3) it increases the chances of compute-data co-location \cite{hdfs-msst10}.
With caching, extra block replicas are created in memory for improving read I/O performance and reducing read latencies \cite{hdfs-caching-webpage}.

In OctopusFS, the blocks are replicated and distributed both across nodes and across storage tiers based on the decision of an automated block placement policy.
For example, a block may have 1 replica in memory, 1 on SSD, and 1 on HDD on three different nodes (see F1B1 in Figure \ref{fig:data-placement}(b)).
Storing data across tiers introduces two additional benefits:
(1) it increases the overall I/O performance and the cluster resource utilization for both write and read operations; and
(2) it enables higher-level systems to make both locality-aware and tier-aware scheduling decisions \cite{octopusfs-sigmod17}.

\subsection{Effect of Tiered Storage in DFSs}
\label{sec:overview:tiering}

In order to study the effects of storing and retrieving block replicas to and from multiple storage tiers, we used the \textit{DFSIO} benchmark \cite{hdfs-msst10}
to write and read 84GB of data in a 12-node cluster with 3 storage tiers: memory, SSD, and HDD. The experimental setup is detailed in Section \ref{sec:evaluation}.
We repeated this experiment in four scenarios:
\begin{itemize}[leftmargin=2em, itemsep=0pt]
	\item \textbf{Original HDFS}, storing 3 replicas on HDDs across 3 different nodes;
	\item \textbf{HDFS with Cache}, storing 1 additional replica in memory on a node that already contains 1 HDD replica;
	\item \textbf{OctopusFS}, which uses a policy to determine the best node and storage tier for storing each of 3 replicas;
	\item \textbf{Octopus++}, our extension of OctopusFS with smart ML-based policies that dynamically move existing replicas between storage tiers.
\end{itemize}

\noindent
The average write and read throughput per node for the four scenarios are shown in Figures \ref{fig:dfsio-thru}(a) and \ref{fig:dfsio-thru}(b), respectively.
We focus on the average throughput per node because the total bandwidth is linear with the number of nodes \cite{hdfs-msst10}.
Comparing the I/O throughput between \textit{Original HDFS} and \textit{OctopusFS} during the generation of the first 42GB of data, we observe that \textit{OctopusFS} achieves a 54\% increase on average write throughput over \textit{Original HDFS} (from 87 to 135MB/s).
Up until that point, \textit{OctopusFS} places the three replicas of each block on the three different tiers -- memory, SSD, and HDD.
Placing replicas on multiple tiers has a modest effect on write performance since the data are written in a pipeline and the performance is bottlenecked by storing 1 replica on HDD.
However, by placing 1 replica in memory and 1 on SSD, the average read throughput increases 3.7x over storing all replicas on HDDs.
Enabling caching in HDFS has no effect on the observed write throughput as caching takes place asynchronously, but
the introduction of 1 cached replica in memory leads to a 2x higher read throughput compared to \textit{Original HDFS}.

The trends in Figures \ref{fig:dfsio-thru}(a) and \ref{fig:dfsio-thru}(b) change after the generation of 42GB of data because the aggregated memory available to the file systems is exhausted.
After that point, \textit{OctopusFS} places either 1 replica on SSD and 2 replicas on HDD or the other way around, reducing the I/O benefits to only 28\% and 36\% of average write and read throughput over \textit{Original HDFS}, respectively.
Enabling caching in HDFS has no effect on write or read throughput anymore as there is no space available to cache any data in memory.

The above results showcase the benefits of tiered storage in distributed file systems in terms of improved I/O performance and better resource utilization.
At the same time, they highlight some important issues from making static data placement decisions. 
First, as time goes by, memory fills up with files that may no longer be needed, which prevents newer or more important files from getting stored there.
On the other hand, some files may be accessed more frequently than others and, hence, it would be more beneficial to move or copy them in memory.
Finally, several previous studies \cite{pacman-nsdi12, mixapart-fast13} have shown that file access patterns evolve over time so a file system must be able to adjust its behavior in order to avoid big variations in performance, as the ones observed in Figure \ref{fig:dfsio-thru}.
Octopus++, which utilizes ML-based policies (see Section \ref{sec:mlapproach}) for automatically moving file replicas up and down the storage tiers, is able to maintain the same performance as OctopusFS for both writes and reads \textit{throughout the entire experiment} (see Figure \ref{fig:dfsio-thru}).

\subsection{Adaptive Tiered Storage Management}
\label{sec:overview:adaptive}

Based on the above results and discussion,
it is essential to build \textit{mechanisms and algorithms for automatically moving data across the storage tiers over time} in order to avoid wasting resources and missing various optimization opportunities.
We have separated the data movement into two categories based on the way the data moves betweens tiers and formulated the following two definitions:

\begin{definition}
\textbf{Replication downgrade} is the process of (i) moving a file replica from a higher storage tier to a lower one, or (ii) deleting a file replica.
\end{definition}

\begin{definition}
\textbf{Replication upgrade} is the process of (i) moving a file replica from a lower storage tier to a higher one, or (ii) creating a new file replica.
\end{definition}

\noindent
We have identified four important decision points that are necessary for guiding replication downgrades or upgrades:
\begin{enumerate}[leftmargin=2em, itemsep=0pt]
	\item When to start the downgrade (upgrade) process
	\item Which file to downgrade (upgrade)
	\item How to downgrade (upgrade) the selected file
	\item When to stop the downgrade (upgrade) process
\end{enumerate}

\noindent
These 4 decision points constitute a \textit{generalization} of both conventional cache management as well as tiering and caching in hierarchical storage management (HSM).
For example, consider a database buffer cache that stores data blocks read from disk.
The ``downgrade'' (i.e., eviction) process starts when the buffer cache gets full (decision \#1).
A data block is selected based on a cache eviction policy such as LRU (decision \#2) and is deleted from the cache (decision \#3).
The eviction policy is invoked again until there is enough room in the buffer cache to fit the new data (decision \#4).
In an HSM system, when the access frequency of a file $f$ becomes higher than a threshold (decision \#1), $f$ (decision \#2) is moved from the HDD layer to the SSD layer (decision \#3).

In a multi-tier DFS, there are multiple interesting options for all four decision points and by treating them differently we get better separation of concerns.
For instance, the system does not need to wait until a storage tier is full to initiate a downgrade process; rather, it can start it proactively in order to overlap the downgrade of a file with the creation of new files.
Similarly, the system does not need to wait until a file is accessed to upgrade it but can start moving it to a higher tier if it expects the file to be used in the near future.
In addition, when the system decides to downgrade or upgrade a file, it then needs to decide whether to delete, move, or copy the file and where.
All these decisions will be handled through \textit{pluggable downgrade and upgrade policies}, elaborated in Sections \ref{sec:mlapproach}-\ref{sec:upgrade}.
The decisions are made at the granularity of files (rather than blocks) since previous work \cite{pacman-nsdi12, big-sql-caching-socc16} has shown that performance improvement is attained only when entire files are present in a higher tier (called the ``all-or-nothing'' property in \cite{pacman-nsdi12}).

\subsection{System Design and Implementation}
\label{sec:overview:impl}

\begin{figure}
	\centering
	\includegraphics[width=0.48\textwidth]{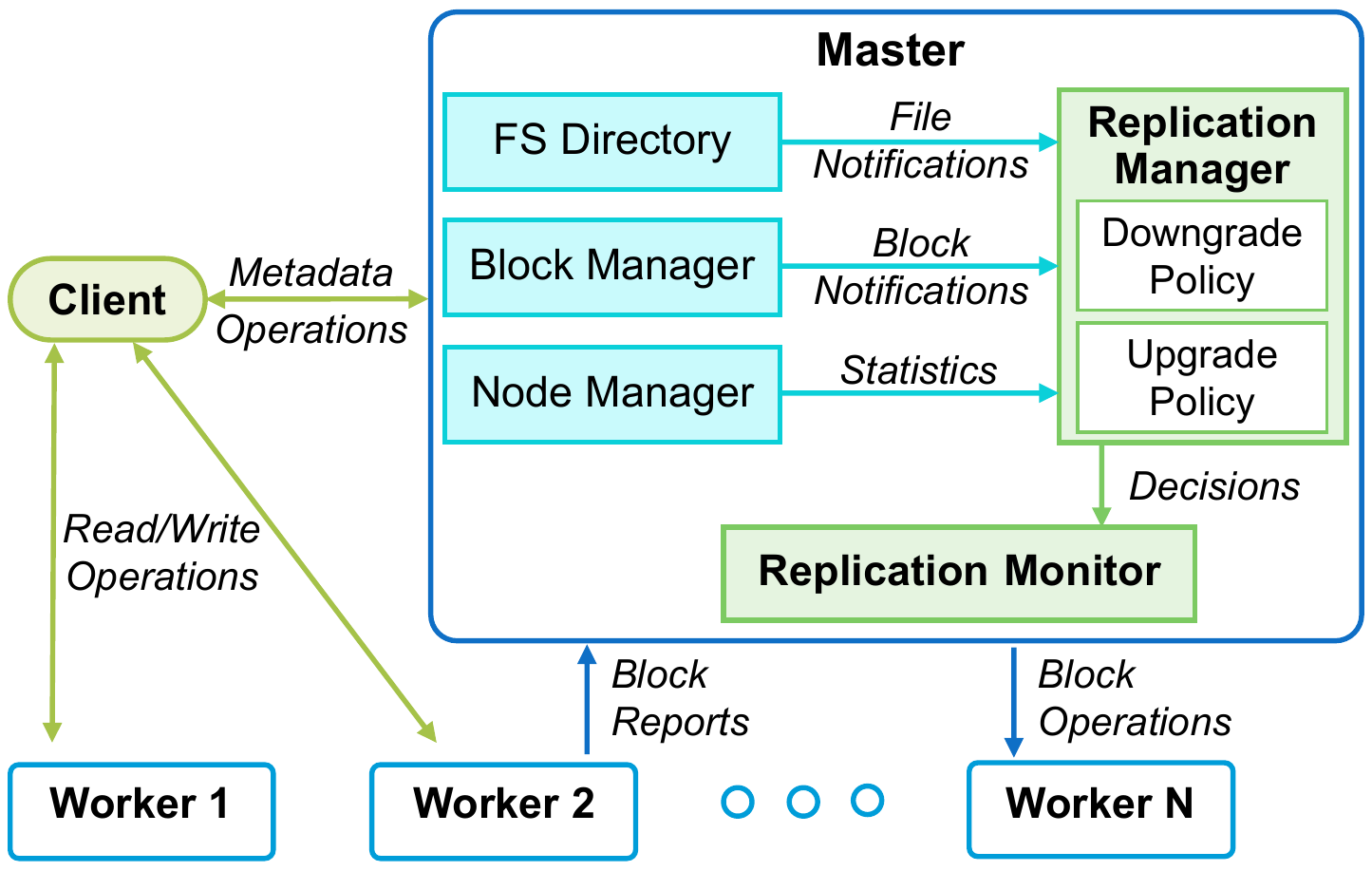}
	\vspace{0.5ex}
	\caption{Octopus++ system architecture}
	\label{fig:system-architecture}
	\vspace{-1ex}
\end{figure}

HDFS, and by extension OctopusFS, uses a multi-master/ worker architecture that consists of \textit{Masters}, \textit{Workers}, and \textit{Clients}.
Each Master contains 
(i) the \textit{FS Directory}, which offers a traditional hierarchical file organization and operations;
(ii) a \textit{Block Manager}, which maintains the mapping from file blocks to nodes and storage tiers; and
(iii) a \textit{Node Manager}, which contains the network topology and maintains node statistics. 
The Workers are responsible for 
(i) storing and managing file blocks on the storage media;
(ii) serving read and write requests from Clients; and
(iii) performing block creation, deletion, and replication upon instructions from the Masters.
The Client exposes APIs for all typical file system operations such as creating and deleting directories or writing and reading files.

We have extended OctopusFS by adding a \textit{Replication Manager} in the Masters
for orchestrating the automatic data movement across the storage tiers, based on the decisions of pluggable downgrade and upgrade policies.
The policies implement 4 main methods that correspond to the 4 core decision points and callback methods for receiving notifications after a file creation, access, modification, or deletion.
Finally, the policies have access to file and node statistics maintained by the system in order to make informed decisions.
In addition, a \textit{Replication Monitor} is responsible for handling the data movement requests from the Replication Manager, as well as monitoring the overall system for any over- or under-replicated blocks.
The Replication Manager and Monitor are the generalized components of AutoCache, our previous framework for admitting and evicting files from the HDFS cache \cite{autocache-icdew19}.
We also modified the Workers to enable the transfer of blocks between storage tiers efficiently.
We did not modify the Client and kept it backward compatible with OctopusFS and HDFS.
For ease of reference, we call our version of the system \textit{Octopus++}.
The policies and their parameters are tunable and their values can be set in the file system's configuration file.
The high-level system architecture of Octopus++ is shown in Figure \ref{fig:system-architecture}.

Even though we implemented our approach in OctopusFS, it is not specific to the internal workings of OctopusFS.
We are confident our framework can be easily implemented in (i) HDFS with caching; (ii) an in-memory distributed file system (e.g., Alluxio, GridGain); or (iii) an HSM system (e.g., ReCA, Hermes) for deciding when and what data to move across the available storage tiers.

\section{File Access Pattern Modeling}
\label{sec:mlapproach}

Previous studies have shown that file access behavior is not random; it is driven by application programs and analytical needs, leading to various types of data re-access patterns \cite{pacman-nsdi12, mr-workloads-vldb12}.
For example, some data may be shared by multiple applications and reused for a few hours before becoming cold, while others are reused for longer periods such as days or weeks.
In addition, data access patterns tend to evolve over time as users and applications are added and removed from a cluster \cite{self-star-icac04}.
The above two observations have motivated our approach of modeling file access patterns and creating a \textit{feature-based classifier} to predict whether a file will be accessed in the near future (and hence, should be upgraded) or it has become cold (and hence, should be downgraded).
The overall approach comprises data preparation, normalization, online incremental training, and binary classification with gradient boosted trees
 (discussed next).

\subsection{Training Data Preparation}
\label{sec:mlapproach:training}

The three most important factors that can influence a replacement or prefetching process in a cache are:
(i) \textit{recency}, i.e., the time of the last file access;
(ii) \textit{frequency}, i.e., the number of accesses to the file; and
(iii) the \textit{size} of the file \cite{web-cache-survey-ijasca11}.
All typical file systems already maintain each file's size, last access time, and creation time.
Even though it would be easy to keep track of access frequencies, that would not reveal any information regarding potential re-access patterns.
Thus, we maintain the last $k$ access times for each file (overheads are discussed in Section \ref{sec:eval:overheads}), which, combined with the file size and creation time, constitute our input data.

\begin{figure}
	\centering
	\includegraphics[width=0.47\textwidth]{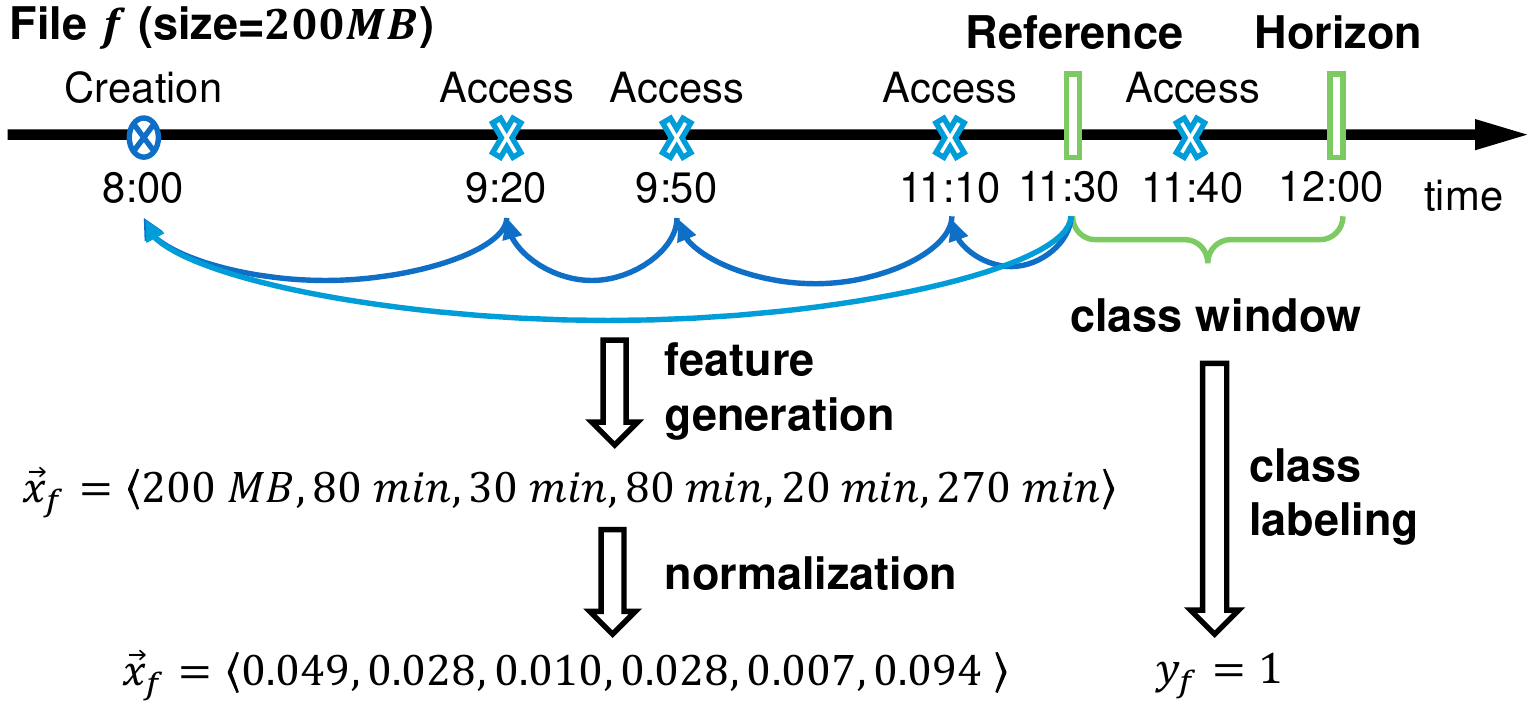}
	\vspace{1.5ex}
	\caption{Pipeline for training data preparation}
	\label{fig:training-data-prep}
	\vspace{-1.5ex}
\end{figure}

The next data preparation step in a classification pipeline is the generation of the \textit{feature vectors} $\vec{x}_i$ and the \textit{class labels} $y_i$, shown in Figure \ref{fig:training-data-prep}.
Timestamps are not good feature candidates for machine learning since their value constantly increases over time.
Hence, we use the timestamps to generate \textit{time deltas}, that is, the time difference between
(i) two consecutive time accesses,
(ii) the oldest time access and creation time,
(iii) a reference time and the most recent access, and
(iv) a reference time and creation time.
A \textit{reference time} is a particular point in time chosen to separate the perceived ``past'' from the perceived ``future''.
The past represents when and how frequently a file has been accessed and, thus, it is used to generate the feature vectors $\vec{x}_i$.
On the other hand, the future shows whether the file will be re-accessed in a given forward-looking \textit{class window}, which is used to generate the class label $y$: if a file is accessed during the window, then $y\!=\!1$; otherwise $y\!=\!0$.
Note that by sliding the reference time in the time axis, we can generate multiple training points (i.e., feature vectors and corresponding class values) based on the access history of a single file.

The final step in our data preparation involves normalizing the features by rescaling all the values to lie between 0 and 1.
Normalization is performed by diving the time deltas by a maximum time interval (e.g., 1 month), which is useful for avoiding outliers from situations where a file was not accessed for a long time.
Figure \ref{fig:training-data-prep} shows a complete example of the training data preparation process.
As the file is accessed $3$ times before the chosen reference time, the feature vector will contain $5$ normalized time deltas as explain above and one file size feature.
The remaining $k-3$ access-based features are encoded as missing values.
The class value is set to 1 since the file is accessed within the class window.

\subsection{Incremental Learning}
\label{sec:mlapproach:incr}

In supervised learning, data $D = ((\vec{x}_1, y_1), (\vec{x}_2, y_2), ..., $ $(\vec{x}_n, y_n))$ consisting of feature vectors $\vec{x}_i$ and class values $y_i$ are used to infer a model $M \approx p(y|\vec{x})$.
Unlike traditional batch (offline) learning, which trains a model from a fixed training dataset $D$, \textit{incremental learning} dynamically refines a model with new data points as they become available gradually over time \cite{incremental-learn-esann16}.
The benefits are that it is unnecessary to determine a fixed training window and that the model naturally adjusts and adapts to changes over time.

The key requirement for incremental learning is the ability to generate training data (i.e., both feature vectors and class values) efficiently while the system is running.
Our data preparation pipeline 
(recall Section \ref{sec:mlapproach:training})
makes this possible as follows.
Suppose our goal is to build a model that predicts whether some file will be re-accessed in a given class window of size $w$ (e.g., in the next 30 minutes).
The system, at the current point in time $t_c$ and for some file $f$, can generate a training data point via:
(1) setting the reference time $t_r = t_c - w$ (i.e., setting the reference time 30 minutes before the current time);
(2) generating the feature vector based on the file size, creation time, and access times before $t_r$ as explained in Section \ref{sec:mlapproach:training}; and
(3) determining the class label based on whether the file was accessed during the time interval between $t_r$ and $t_c$.
By repeating the above three steps periodically for a sample of (or all) the files in the file system, we can generate continuous data points for (re-) training the model over time.
Note that we can efficiently compute the feature vectors for each file incrementally since time deltas between consecutive file accesses do not change over time.
Finally, in order to ensure the generation of positive training data (i.e., data points with class value $y=1$), we also repeat the above three steps right after a file is accessed (but only for that file).

\subsection{Learning Model Selection}
\label{sec:mlapproach:model}

The following requirements for a machine learning model are desirable, since we plan to use the model in a real system:
\begin{itemize}[leftmargin=1em]
	\itemsep0.1em 
	\item \textbf{Accurate:} The model must be able to accurately predict whether a file will be accessed soon or not be accessed for some time.
	\item \textbf{Efficient:} Both model training and predictions must be inexpensive in terms of computational and storage requirements.
	\item \textbf{Adaptable:} The model must support incremental learning and efficiently adapt to new workloads patterns.
\end{itemize}

\noindent
As a learning model we selected \textit{XGBoost} \cite{xgboost-kdd16}, a state-of-the-art gradient boosting tree algorithm that satisfies all three requirements.
The XGBoost model has the form of an ensemble of weak models (single trees), which is trained following a stage-wise procedure under the same (differentiable) loss function \cite{xgboost-kdd16}.
XGBoost is very effective in practice and has been used in multiple winning solutions for regression, classification, and ranking in recent ML competitions \cite{gradient-boosting-blog}.
As an ensemble model, XGBoost is more complex and less understandable than simple regression or rule-based models.
In response, various methods have been proposed recently to help admins compute feature importance measures and understand the reasons of individual predictions \cite{interpreting-ml-nips2017, explaining-ml-kdd2016}.

We also considered other well-known classifiers such as Naive Bayes, Bayesian Belief Networks (BBN), Support Vector Machines (SVM), and Artificial Neural Networks (ANN), but each failed to satisfy some of our needs.
Naive Bayes assumes that attributes are conditionally independent of one another, and thus cannot be used effectively to learn a sequence of access patterns.
BBNs model attribute dependence in networks, but require a priori knowledge about the structure of the network, which is exactly what we are trying to determine \cite{self-star-icac04}.
Finally, we empirically found SVM and ANN to have a much higher cost in terms of training time (up to two orders of magnitude slower compared to XGBoost) and lower accuracy than XGBoost.
On the contrary, \textit{XGBoost requires minimal storage, is fast to train and make predictions, and can learn incrementally over time}.

\vspace{2pt}
\noindent
\textbf{Hyperparameter tuning:}
We used XGBoost2 v0.60 and set the learning objective to be logistic regression for binary classification.
In pre-analysis, we found that only two configuration parameters had a noticeable effect on the XGBoost performance: the maximum depth $d$ of the trees and the number of rounds $r$ for boosting \cite{xgboost-docs}.
Hence, we used grid search to optimize them offline using training data from our two workload traces (see Section \ref{sec:evaluation}) and found the same values for both workloads: $d=20$ and $r=10$.
We used default values for other configuration parameters.

\subsection{File Access Predictions}
\label{sec:mlapproach:predict}

Section \ref{sec:mlapproach:incr} described how a model $M$ is trained incrementally over time.
Before the system can start using $M$ to make predictions, it needs to ensure that $M$ has been trained enough.
To achieve this, the system will occasionally use some training data points for evaluating the performance of $M$ (before using them for training $M$).
When the classification error rate drops below a certain threshold (e.g., 0.01), then the system can start using $M$ for predicting whether a file will be accessed in the forward-looking class window of size $w$ (e.g., in the next 30 minutes).
At the current point in time $t_c$ and for some file $f$, the system can get the prediction via:
(1) setting the reference time $t_r = t_c$ 
(i.e., setting the reference time equal to the current time);
(2) generating the feature vector
based on the file size, creation time, and access times before $t_r$ 
as explained in Section \ref{sec:mlapproach:training}; and
(3) using the model to predict the class label of $f$.
Specifically, an XGBoost model will return a \textit{probability score} indicating how likely is $f$ to be accessed in the next $w$ minutes.

The probability score is then used by the policies to decide which file(s) to upgrade or downgrade.
We generate two separate models for this purpose --- one for the upgrade policy and one for the downgrade policy --- whose only difference lies in the class window size $w$.
The upgrade policy wants to determine which files will be accessed in the immediate future and, hence, we want to set a small $w$ (e.g., 30 minutes).
On the other hand, the downgrade policy wants to determine which files have become cold, that is, will not be accessed for some time; hence, we want to set a large $w$ (e.g., 6 hours).
The two policies are further elaborated in Sections \ref{sec:downgrade} and \ref{sec:upgrade}.

\section{Downgrade Policies}
\label{sec:downgrade}

Algorithm \ref{alg:downgrade} shows the outline of the downgrade process  responsible for moving a file replica from a higher storage tier to a lower one.
The procedure utilizes the four main methods implemented by a downgrade policy, which correspond to the four decision points discussed in Section \ref{sec:overview:adaptive}.
The downgrade procedure is invoked every time some data is added to a storage tier (e.g., after a file creation or replication) but the actual process starts based on the policy's decision (line \#2).
At that point, the policy is responsible for selecting a file to downgrade (line \#4) and the target storage tier (line \#5).
The system will then schedule the downgrade request (line \#6), which will take place asynchronously.
Finally, the process will repeat until the policy decides to stop it (line \#7).

\subsection{When to Start the Downgrade Process}
\label{sec:downgrade:start}

In a traditional cache, an object is discarded when the cache is full and a new object needs to enter the cache.
This approach ensures that the cache is fully utilized and works well for fixed-size disk pages or small web objects.
However, typical file sizes in analytics clusters are in the order of tens to hundreds of MBs \cite{pacman-nsdi12, mr-workloads-vldb12}, so having a file write wait for other files to be downgraded would introduce significant latency delays.
Hence, it is crucial for the system to be proactive and for the downgrade process from a tier $T$ to start before $T$ is full.
All of our policies will start the downgrade process from a tier when its used capacity becomes greater than a threshold value (e.g., 90\%), allowing for a more efficient overlapping between file writes and file downgrades.
In other words, this threshold constitutes a tradeoff between being proactive and utilizing a (high) storage tier as much as possible.

\begin{algorithm}[t]
	\small
	\renewcommand\algorithmicindent{0.9em}
	\caption{Downgrade process outline}
	\label{alg:downgrade}
	\begin{algorithmic}[1]
		\Procedure{Downgrade}{$\mathit{StorageTier~fromTier}$}
		\If{$\mathit{policy}.\mathbf{startDowngrade}(fromTier)$}
		\Repeat
		\State $\mathit{file} = \mathit{policy}.\mathbf{selectFileToDowngrade}(\mathit{fromTier})$
		\State $\mathit{toTier} = \mathit{policy}.\mathbf{selectDowngradeTier}(\mathit{file, fromTier})$  
		\State $\mathit{downgradeFile}(\mathit{file, fromTier, toTier})$
		\Until{$\mathit{policy}.\mathbf{stopDowngrade}(fromTier)$}
		\EndIf
		\EndProcedure
	\end{algorithmic}
\end{algorithm}

\begin{table*}
	\centering
	\caption{Downgrade policies}
	\vspace{1ex}
	\renewcommand{\arraystretch}{1.1}
	\small
	\begin{tabular}{| l | l | l |}
		\hline
		{\bf Acronym} & {\bf Policy Name} & {\bf Description} \\ 
		\hline
		LRU		&	Least Recently Used	& 	Downgrade the file that was accessed less recently than any other \\
		LFU		&	Least Frequently Used	& 	Downgrade the file that was used least often than any other \\
		LRFU  	&	Least Recently \& Frequently Used	& 	Downgrade the file with the lowest weight based on recency and frequency \\
		LIFE		&	LIFE (PACMan \cite{pacman-nsdi12})	&	Downgrade either the old LFU file or the largest file \\
		LFU-F	&	LFU-F (PACMan \cite{pacman-nsdi12})	& 	Downgrade the LFU file among files older than a time window \\
		EXD		&	Exponential Decay (Big SQL \cite{big-sql-caching-socc16})	& 	Downgrade the file with the lowest weight based on recency and frequency \\
		XGB		&	XGBoost-based Modeling	& 	Downgrade the file with the lowest access probability in the distant future \\
		\hline
	\end{tabular}
	\label{table:downgrade-policies}
	\vspace{-1.3ex}
\end{table*}

\subsection{Which file to downgrade}
\label{sec:downgrade:file}

Once the downgrade process is activated for a particular storage tier $T$, the policy must select a file to remove from $T$ in order to make room for new data.
This particular decision is known as the \textit{replacement or eviction policy} in the literature with a long history of related work \cite{web-cache-survey-ijasca11, survey-cache-repl-csur03}.
For comparison purposes, we have implemented three conventional eviction policies, three related policies from recent literature, and one new policy, listed in Table \ref{table:downgrade-policies}.

\vspace{2pt}
\noindent \textbf{LRU (Least Recently Used)}
selects the file used least recently.
LRU is widely used and is designed to take advantage of the temporal locality often exhibited in data accesses.

\vspace{2pt}
\noindent \textbf{LFU (Least Frequently Used)}
selects the file with the least number of accesses.
LFU is a typical web caching policy that keeps more popular files and evicts rarely used ones.

\vspace{2pt}
\noindent \textbf{LRFU (Least Recently \& Frequently Used)}
selects the file with the lowest weight, which is computed for each file based on both the recency and frequency of accesses.
The weight $W$ for a file $f$ is initialized to 1 when $f$ is created and updated each time $f$ is accessed based on Formula \ref{eq:lrfu}:
\begin{equation}\label{eq:lrfu}
W = 1 + \frac{H * W }{(timeNow - timeLastAccess) + H}
\end{equation}
Parameter $H$ represents the ``half life'' of $W$, i.e., after how much time the weight is halved.
For e.g., if $H=6$ hours and a file is accessed 6 hours after its last access, then its new weight will equal 1 plus half the old weight.
Hence, files that are recently accessed multiple times will have a large weight, as opposed to files accessed a few times in the past.

\vspace{2pt}
\noindent \textbf{LIFE}
aims at minimizing the average completion time of jobs in PACMan \cite{pacman-nsdi12} by prioritizing small and recent files.
Specifically, LIFE divides the files into two partitions: $P_{old}$ containing the files that have not been accessed for at least some time window (e.g., 9 hours) and  $P_{new}$ with the rest.
If $P_{old}$ is not empty, then the LFU file is selected from it.
Otherwise, LIFE selects the largest file from $P_{new}$.

\vspace{2pt}
\noindent \textbf{LFU-F}
aims at maximizing cluster efficiency in PACMan \cite{pacman-nsdi12} by evicting less frequently accessed files.
LFU-F divides the files in the same two partitions as LIFE, namely, $P_{old}$ and $P_{new}$.
If $P_{old}$ is not empty, then the LFU file is selected from it.
Otherwise, LFU-F selects the LFU file from $P_{new}$.

\vspace{2pt}
\noindent \textbf{EXD (Exponential Decay)}
explores the tradeoff between recency and frequency in data accesses in Big SQL \cite{big-sql-caching-socc16}.
In particular, it selects the file with the lowest weight $W$ computed using the following formula:
\begin{equation}\label{eq:exd}
W = 1 + W * e^{-\alpha ~*~ (timeNow ~-~ timeLastAccess)}
\end{equation}
The parameter $\alpha$ determines the weight of frequency vs. recency and it is set to $1.16*10^{-8}$ based on \cite{big-sql-caching-socc16}.

\vspace{2pt}
\noindent \textbf{XGB (XGBoost-based Modeling)}
incrementally trains and utilizes an XGBoost model (recall Section \ref{sec:mlapproach}) for deciding which file will not be accessed in the distant future.
Specifically, XGB will compute the access probability for the $k$ least recently used files and select the file with the lowest access probability to downgrade.
We compute probabilities for LRU files in order to avoid cache pollution with files that are never evicted, while we limit the computations to $k$ files in order to bound the (low) overhead of building the features and using the model.
In practice, we set $k$ to be large (e.g., $k=200$), and it has had limited impact on our workloads.

\subsection{How to downgrade the selected file}
\label{sec:downgrade:tier}

When an object is selected for cache eviction, it is typically simply deleted.
In our case, however, when a file is selected for downgrade from a higher tier, we typically want to move that file replica to a lower tier in order to retain the same number of replicas, and hence, maintain the file system's properties of high fault tolerance and availability.
This decision entails selecting one lower storage tier for moving the file replica.
For this purpose, we adapted the data placement policy used by OctopusFS, which makes decisions by solving a \textit{multi-objective optimization problem}.
In particular, the policy aims at finding a Pareto optimal solution that optimizes 4 objectives simultaneously:
(1) fault tolerance for avoiding data loss due to failures;
(2) load balancing for distributing I/O requests across storage tiers;
(3) data balancing for distributing data blocks across storage tiers; and
(4) throughput maximization for optimizing the overall I/O throughput of the cluster.
For details, see \cite{octopusfs-sigmod17}.

\subsection{When to stop the downgrade process}
\label{sec:downgrade:stop}

In conventional caches, eviction stops when there is enough room in the cache to fit the newly inserted object.
Since we start the downgrade process proactively (i.e., before the cache is full), we need a different approach for stopping it.
Specifically, all policies will stop the downgrade process from a storage tier $T$ when its used capacity becomes lower than a threshold value (e.g., 85\%), allowing for a small percent of $T$'s capacity to be freed together while still maintaining its high utilization.

\setlength{\textfloatsep}{\oldtextfloatsep}

\section{Upgrade Policies}
\label{sec:upgrade}

\begin{algorithm}[t]
	\small
	\renewcommand\algorithmicindent{1em}
	\caption{Upgrade process outline}
	\label{alg:upgrade}
	\begin{algorithmic}[1]
		\Procedure{Upgrade}{$\mathit{StorageTier\,fromTier,File\,accessedFile}$}
		\If{$\mathit{policy}.\mathbf{startUpgrade}(fromTier, accessedFile)$}
		\Repeat
		\State $\mathit{file} = \mathit{policy}.\mathbf{selectFileToUpgrade}(\mathit{fromTier})$
		\State $\mathit{toTier} = \mathit{policy}.\mathbf{selectUpgradeTier}(\mathit{file, fromTier})$  
		\State $\mathit{upgradeFile}(\mathit{file, fromTier, toTier})$
		\Until{$\mathit{policy}.\mathbf{stopUpgrade}(fromTier)$}
		\EndIf
		\EndProcedure
	\end{algorithmic}
\end{algorithm}

\begin{table*}
	\centering
	\caption{Upgrade policies}
	\vspace{1ex}
	\renewcommand{\arraystretch}{1.1}
	\small
	\begin{tabular}{| l | l | l |}
		\hline
		{\bf Acronym} & {\bf Policy Name} & {\bf Description} \\ 
		\hline
		OSA	&	On Single Access	& 	Upgrade a file into memory upon access (if not there already) \\
		LRFU  	&	Least Recently \& Frequently Used	& 	Upgrade a file if its weight is higher than a threshold \\
		EXD		&	Exponential Decay (Big SQL \cite{big-sql-caching-socc16})	& 	Upgrade a file if its weight is higher than the weight of to-be-evicted files \\
		XGB		&	XGBoost-based Modeling	& 	Upgrade files with high access probability in the near future \\
		\hline
	\end{tabular}
	\label{table:upgrade-policies}
	\vspace{-1ex}
\end{table*}

Algorithm \ref{alg:upgrade} outlines the upgrade process guided by the 4 decision points presented earlier in Section \ref{sec:overview:adaptive}.
The upgrade procedure is invoked (i) every time a file is accessed (but before it is actually read) and (ii) periodically in case the policy wants to make a proactive decision (an accessed file is not available in this case).
Given a storage tier and the file that was just accessed (optional), the upgrade policy is responsible for deciding when the process starts (line \#2), which file to upgrade (line \#4), and the target storage tier (line \#5).
The system will then schedule the upgrade request (line \#6), which will be piggybacked during the subsequent read or take place asynchronously.
Finally, the process will repeat until the policy decides to stop it (line \#7).

\subsection{When to Start the Upgrade Process}
\label{sec:upgrade:start}

Typically, all data accesses in a system that uses a cache must go through the cache first.
If the accessed object $O$ is located in the cache, it will be served from there; otherwise, $O$ will be inserted into the cache.
Unlike cache eviction policies, \textit{cache admission} policies are not very common as they complicate the read process without major benefits in a traditional cache \cite{big-sql-caching-socc16}.
In our case, moving a file into a higher storage tier is costlier as it may be executed asynchronously and it may involve a large amount of data (10s to 100s of MBs).
Hence, the decisions of when and what to upgrade are as important as when and what to downgrade.
For comparison purposes, we have implemented two conventional admission policies, one related policy from recent literature, and one new policy, listed in Table \ref{table:upgrade-policies}.

\vspace{2pt}
\noindent \textbf{OSA (On Single Access)}
implements the common approach of upgrading each file when it is accessed and not already present in the memory tier.
With OSA, we do not allow upgrades from the HDD to the SSD tier to avoid 
(i) the overhead associated with moving large amounts of data between disks, and
(ii) under-utilizing the HDDs available in the cluster.

\begin{table*}
	\setlength{\tabcolsep}{6pt}
	\centering
	\caption{Job size distributions. The jobs are binned by their data sizes in our FB and CMU workloads}
	\vspace{1ex}
	\renewcommand{\arraystretch}{1.1}
	\small
	\begin{tabular}{| c | c | r | r | r | r | r | r | r | r |}
		\hline
		Bin	&	Data size	& \multicolumn{2}{| c |}{\% of Jobs}	&		\multicolumn{2}{| c |}{\% of Resources}	&	\multicolumn{2}{| c |}{\% of I/O}	&	\multicolumn{2}{| c |}{Task Time (mins)}	\\
		\cline{3-10}
		&		&	\multicolumn{1}{|c|}{FB}	&	\multicolumn{1}{|c|}{CMU}	&	\multicolumn{1}{|c|}{FB}	&	\multicolumn{1}{|c|}{CMU}	&	\multicolumn{1}{|c|}{FB}	&	\multicolumn{1}{|c|}{CMU}	&	\multicolumn{1}{|p{1cm}|}{\centering FB}	&	\multicolumn{1}{|c|}{CMU}	\\
		\hline
		A	&	0-128MB	&	74.4\%	&	63.4\%	&	25.0\%	&	32.3\%	&	3.2\%	&	10.9\%	&	76.7	&	119.5	\\
		B	&	128-512MB	&	16.2\%	&	29.1\%	&	12.2\%	&	27.9\%	&	16.1\%	&	30.5\%	&	37.6	&	103.2	\\
		C	&	0.5-1GB	&	4.0\%	&	0.9\%	&	7.3\%	&	1.3\%	&	12.0\%	&	2.4\%	&	22.3	&	5.0	\\
		D	&	1-2GB	&	3.0\%	&	4.9\%	&	13.4\%	&	21.0\%	&	19.3\%	&	23.3\%	&	41.0	&	77.6	\\
		E	&	2-5GB	&	1.6\%	&	1.5\%	&	20.8\%	&	15.1\%	&	21.9\%	&	27.8\%	&	63.9	&	55.7	\\
		F	&	5-10GB	&	0.8\%	&	0.3\%	&	21.4\%	&	2.5\%	&	27.5\%	&	5.2\%	&	65.6	&	9.2	\\
		\hline
	\end{tabular}
	\vspace{-1ex}
	\label{table:bins}
\end{table*}

\vspace{2pt}
\noindent \textbf{LRFU (Least Recently \& Frequently Used)}
starts the upgrade process for an accessed file $f$ when the computed weight for $f$ is greater than a threshold value.
The weight takes into account both the recency and the frequency of accesses and is computed using Formula \ref{eq:lrfu}.
The threshold value is empirically set to 3 in order to favor files that are accessed recently multiple times.

\vspace{2pt}
\noindent \textbf{EXD (Exponential Decay)}
is used in Big SQL \cite{big-sql-caching-socc16} for selecting which files to insert into the cache.
If there is enough space in a higher storage tier to fit the accessed file $f$, then $f$ will get upgraded.
Otherwise, EXD will upgrade $f$ only if its weight (computed using Formula \ref{eq:exd}) is higher than the sum of weights of the files that will need to be downgraded to make room for $f$.

\vspace{2pt}
\noindent \textbf{XGB (XGBoost-based Modeling)}
incrementally trains and utilizes an XGBoost model (recall Section \ref{sec:mlapproach}) for predicting if a file will get accessed in the near future.
Specifically, XGB will compute the access probability for the $k$ (e.g., $k=200$) most recently used files and start the upgrade process if the access probability of a file is higher than the discrimination threshold.
In binary classification, the discrimination threshold determines the boundary between the two classes, and it is empirically set to 0.5 (see Section \ref{sec:eval:xgboost}).

\begin{figure*}[t!]
	\centering
	\includegraphics[width=0.325\textwidth]{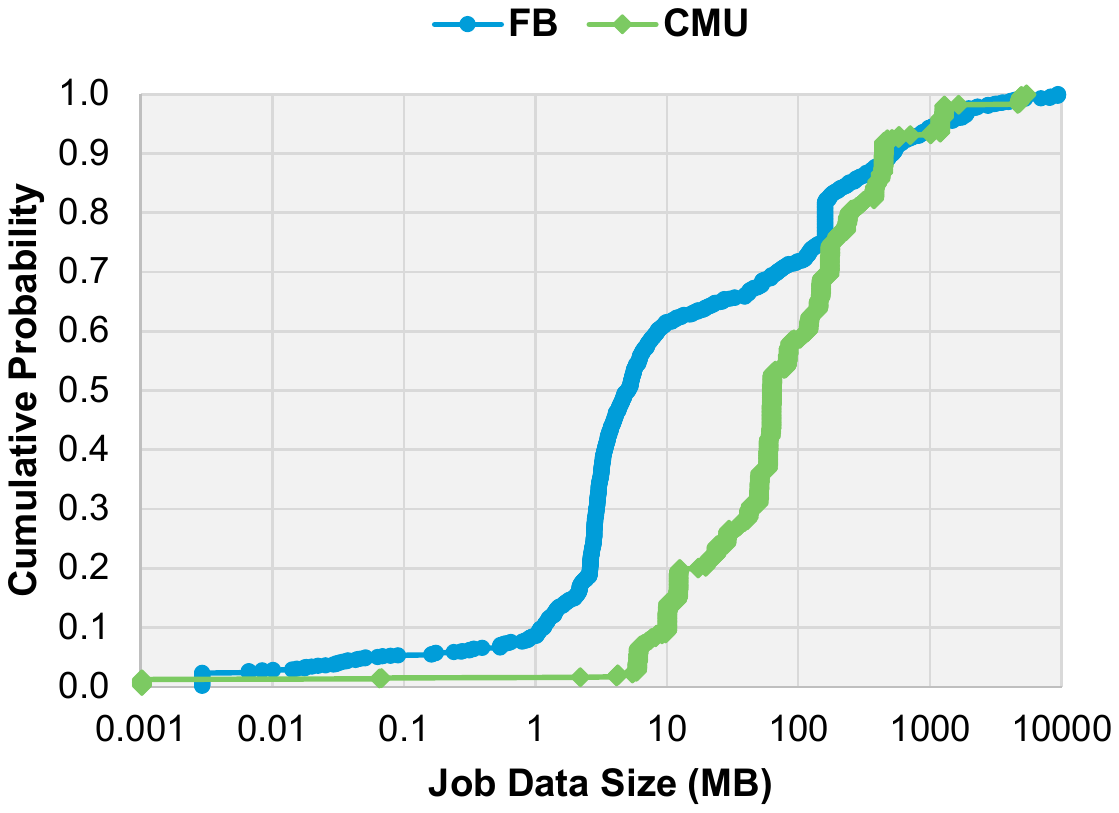}
	\hfil
	\includegraphics[width=0.325\textwidth]{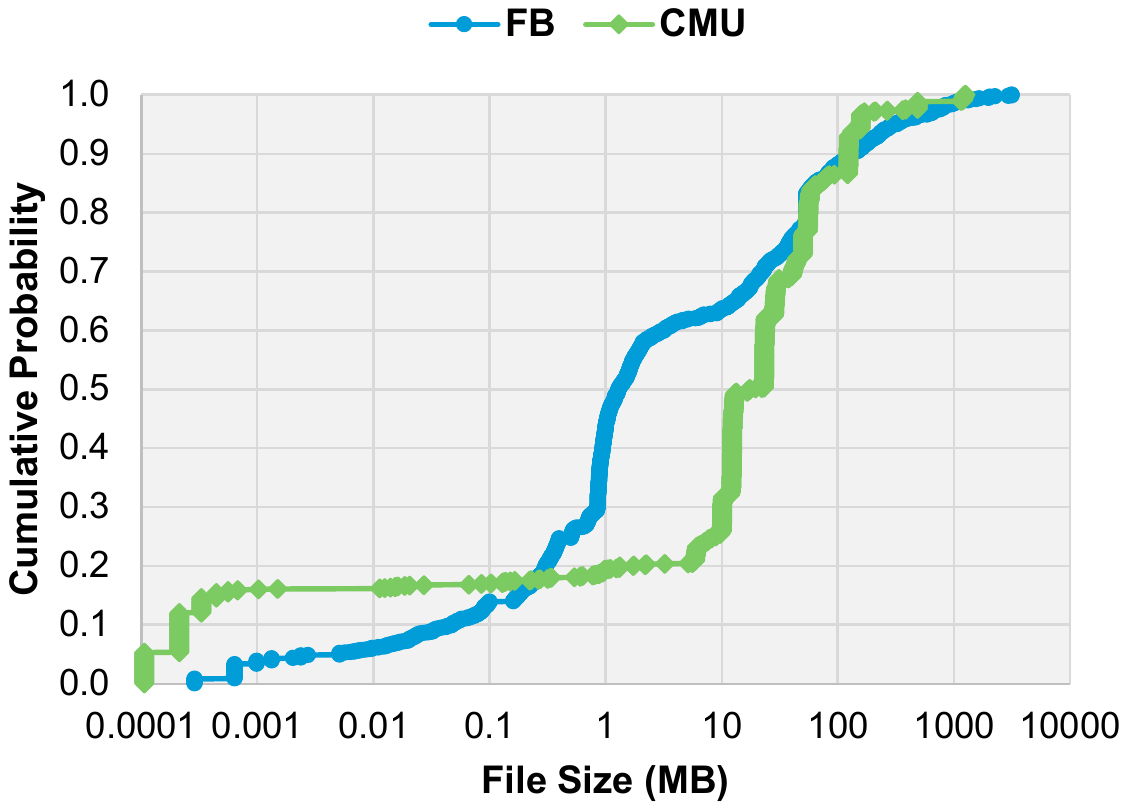}
	\hfil
	\includegraphics[width=0.325\textwidth]{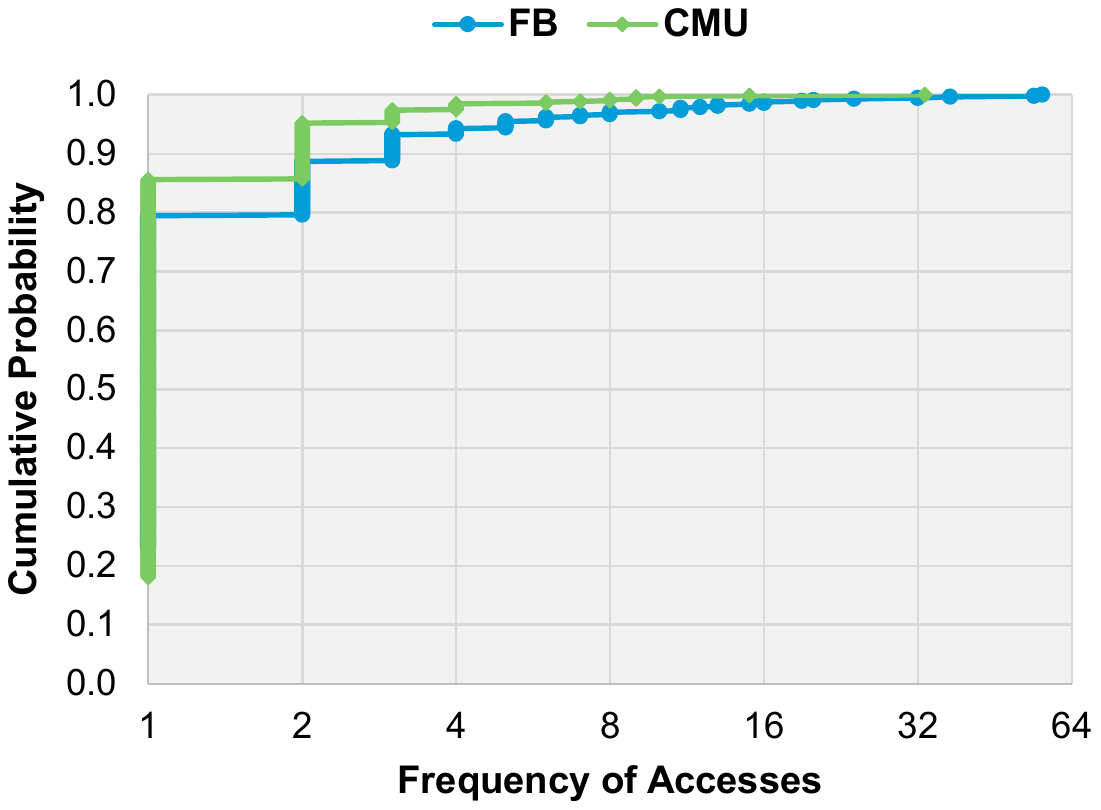}
	\vspace{1ex}
	\caption{Cumulative Distribution Functions (CDFs) for (a) the job data size, (b) the file size, and (c) the frequency of accesses for the FB and CMU workloads. Note the logarithmic x-axis}
	\label{fig:cdf-graphs}
	\vspace{-1ex}
\end{figure*}

\subsection{Which file to upgrade}
\label{sec:upgrade:file}

The decisions of when to start upgrading and which file to upgrade are tightly coupled in the upgrade process.
Hence, all policies will select the file that triggered the process (described in Section \ref{sec:upgrade:start} above) as the file to upgrade.

\subsection{How to upgrade the selected file}
\label{sec:upgrade:tier}

We use the same multi-objective optimization problem formulation with the downgrade policies (recall Section \ref{sec:downgrade:tier}) for selecting a higher storage tier for performing the upgrade, 
while considering
the tradeoffs between fault tolerance, data and load balancing, and throughput maximization.

\subsection{When to stop the upgrade process}
\label{sec:upgrade:stop}

With the exception of XGB, all other policies base their decision to start the upgrade process on the currently accessed file $f$.
If they decide to start, $f$ will be upgraded and the loop terminates.
XGB, on the other hand, will continue the upgrade process until either there are no more files that are likely to be accessed in the near future, or until the total size of the scheduled upgrades exceeds a threshold (e.g., 1GB) to avoid upgrading a large amount of data at once.

\setlength{\textfloatsep}{\oldtextfloatsep}

\section{Experimental Evaluation}
\label{sec:evaluation}

The evaluation is conducted on a 12-node cluster running CentOS Linux 7.2 with 1 Master and 11 Workers.
The Master node has a 64-bit, 8-core, 3.2GHz CPU, 64GB RAM, and a 2.1TB RAID 5 storage configuration.
Each Worker node has a 64-bit, 8-core, 2.4GHz CPU, 24GB RAM, one 120GB SATA SSD, and three 500GB SAS HDDs.
The file systems are configured to use three storage tiers consisting of 4GB of memory, 64GB of SSD, and 400GB of HDD space each for storing file blocks on each Worker node.
The default replication factor is 3 and the block size is 128MB.

\subsection{Workload Properties}
\label{sec:eval:workloads}

\begin{figure*}[t!]
	\centering
	\includegraphics[width=0.47\textwidth]{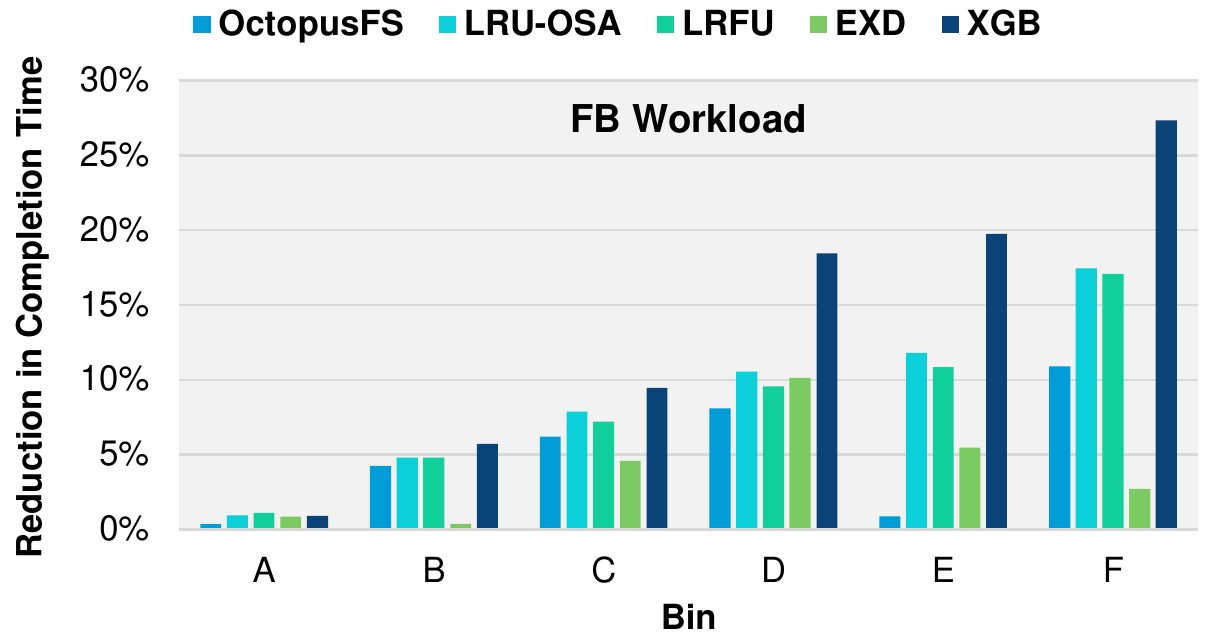}
	\hfil
	\includegraphics[width=0.47\textwidth]{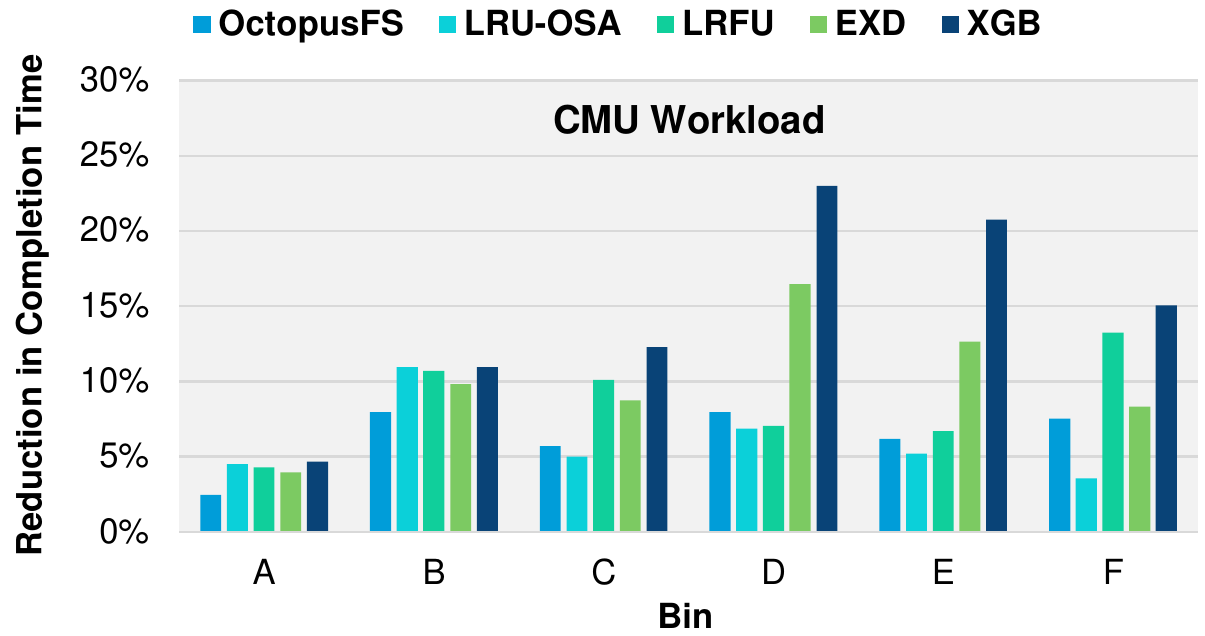}
	\vspace{1ex}
	\caption{Percent reduction in completion time over HDFS for the (a) FB and (b) CMU workloads}
	\label{fig:completion-time}
	\vspace{-1ex}
\end{figure*}

\begin{figure*}[t!]
	\centering
	\includegraphics[width=0.47\textwidth]{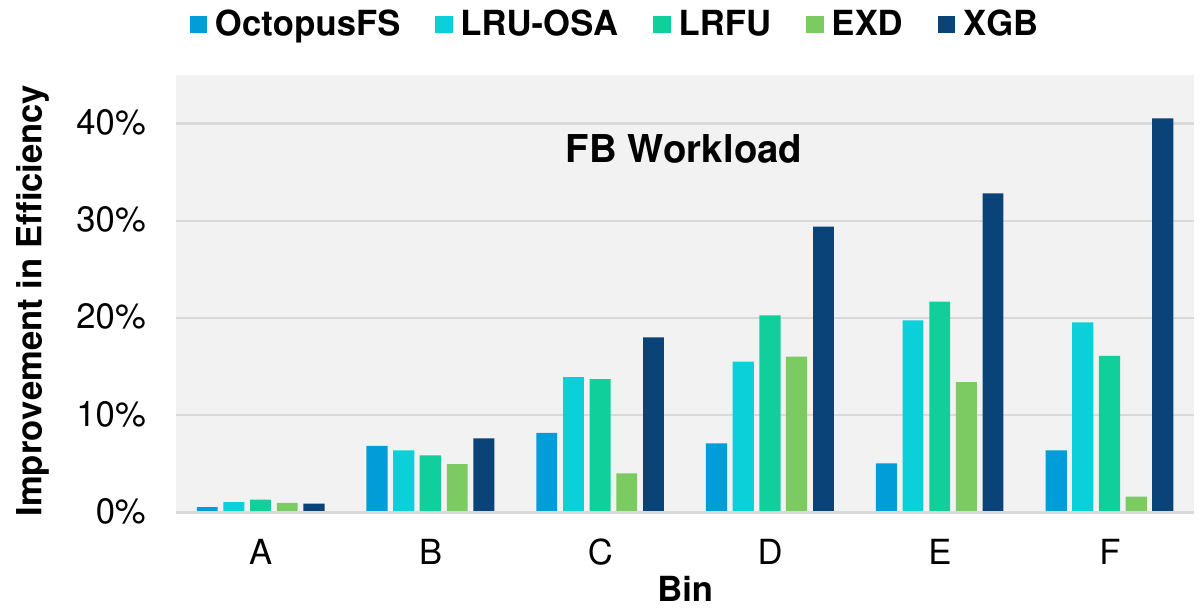}
	\hfil
	\includegraphics[width=0.47\textwidth]{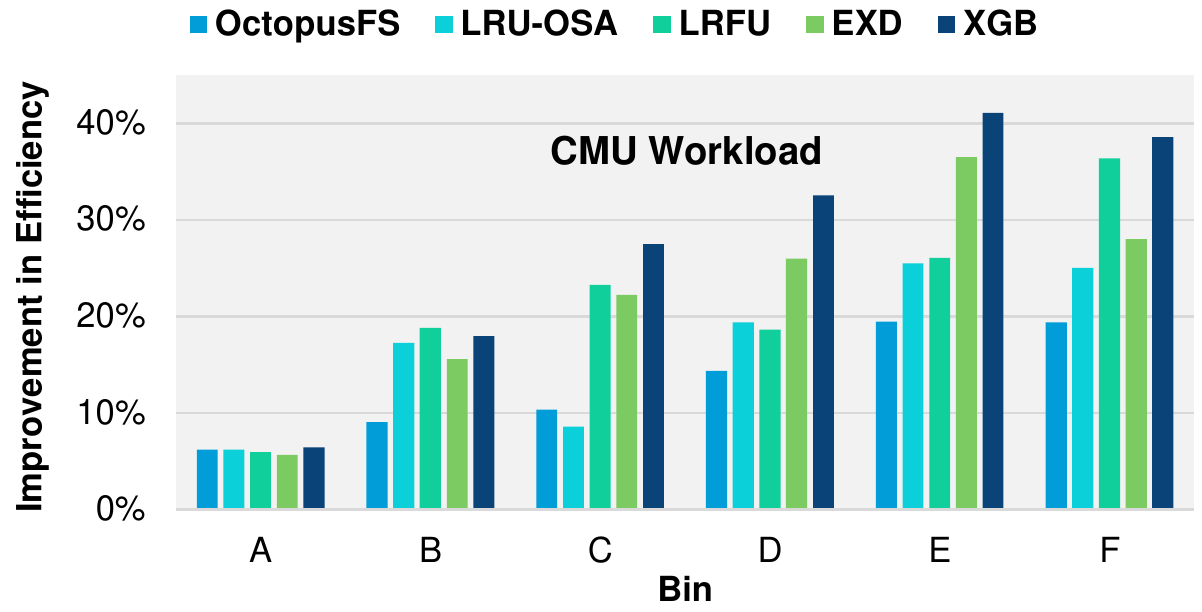}
	\vspace{1ex}
	\caption{Percent improvement in cluster efficiency over HDFS for the (a) FB and (b) CMU workloads}
	\label{fig:cluster-efficiency}
	\vspace{-1.4ex}
\end{figure*}

Our evaluation is based on two workloads derived from real-world production traces from Facebook and Carnegie Mellon University clusters.
The \textit{FB} trace was collected over a period of 6 months from a 600-node Hadoop cluster at Facebook and contains arrival times, durations, file sizes, and other data about executed MapReduce jobs \cite{swim-paper-mascots2011}.
The \textit{CMU} trace contains similar data from scientific MapReduce workloads executed over a period of 31 months at OpenCloud, a 64-node Hadoop cluster \cite{cmu-opencloud-webpage}.
From the traces, we used SWIM \cite{swim-webpage}, a statistical workload injector for MapReduce, to generate two realistic and representative workloads that preserve the original workload characteristics such as the distribution of input sizes and the skewed popularity of data \cite{pacman-nsdi12}.
Using SWIM, we replay each workload with the same inter-arrival times and input/output files as in the original workload, 
allowing us to mimic the access patterns of the files.
In order to reflect the smaller size of our cluster and to simulate the load experienced by the original clusters, we scale down the file sizes proportionately \cite{pacman-nsdi12}.

The derived FB and CMU workloads consist of 1000 and 800 jobs, respectively, scheduled for execution over a 6-hour period.
To separate the effect of storage tiering on different jobs, we split them based on their input data size into 6 bins.
Table \ref{table:bins} shows the distribution of jobs by count, cluster resources they consume, amount of I/O they generate, and aggregate task execution time.
The jobs in both workloads exhibit a heavy-tailed distribution of input sizes, also noted in previous studies \cite{pacman-nsdi12, mr-workloads-vldb12}.
In particular, the FB workload is dominated by small jobs; 74.4\% of them process $<$128MB of data.
However, these jobs only account for 25\% of the cluster resources consumed and perform only 3.2\% of the overall I/O.
On the contrary, FB jobs processing $>$1GB of data, account for 54\% of resources and 68\% of I/O.
The distribution of input sizes is less skewed for CMU with 63.4\% of the jobs processing $<$128MB of data.
Similarly, CMU jobs processing $>$1GB of data, account for 38\% of resources and 56\% of I/O.
In both workloads, even though larger jobs (Bins D-F) constitute only a small fraction of the workload, they account for about half the total task execution time.

In terms of files, the FB and CMU workloads process 1380 and 1305 files with a total size of 92GB and 85GB, respectively.
The popularity of files is also skewed in data-intensive workloads, with a small fraction of the files accessed very frequently, while the rest are accessed less frequently \cite{pacman-nsdi12, mr-workloads-vldb12}.
Specifically, 5.7\% of FB files and 2.8\% of CMU files are accessed more than 5 times.
Such repeatability must be exploited to improve job performance by ensuring their inputs have replicas residing in the highest storage tier (i.e., memory).
In addition, a sizable fraction of the files (23\% for FB and 18\% for CMU) are created but not accessed afterwards.
Hence, it is important for a downgrade policy to identify such cases and remove their replicas from memory early on.
CDFs for the workload statistics are shown in Figure \ref{fig:cdf-graphs}.

\subsection{End-to-End Evaluation}
\label{sec:eval:endtoend}

For this evaluation, we executed the two workloads over HDFS v2.7.7, the default OctopusFS (i.e., without any downgrade or upgrade policies), Octopus++ using the LRU downgrade policy and the OSA upgrade policy (as a baseline for Octopus++), and Octopus++ using all common downgrade and upgrade policies (i.e., LRFU, EXD, and XGB; recall Tables \ref{table:downgrade-policies} and \ref{table:upgrade-policies}).
We compare them using two complementary performance metrics:
(i) the \textit{average completion time} of jobs, and
(ii) the \textit{cluster efficiency} (defined as finishing the jobs by using the least amount of resources \cite{pacman-nsdi12}).

\begin{figure*}[t!]
	\centering
	\subfloat[FB Workload]{
		\includegraphics[width=0.47\textwidth]{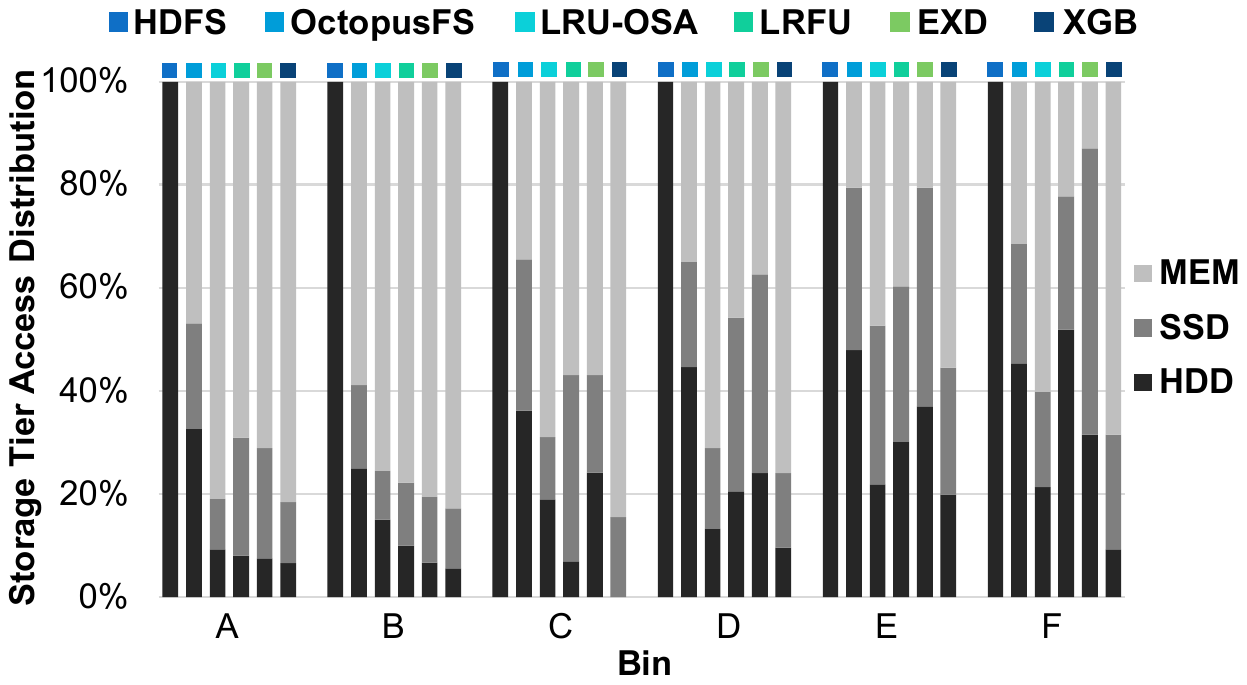}}
	\hfil
	\subfloat[CMU Workload]{
		\includegraphics[width=0.47\textwidth]{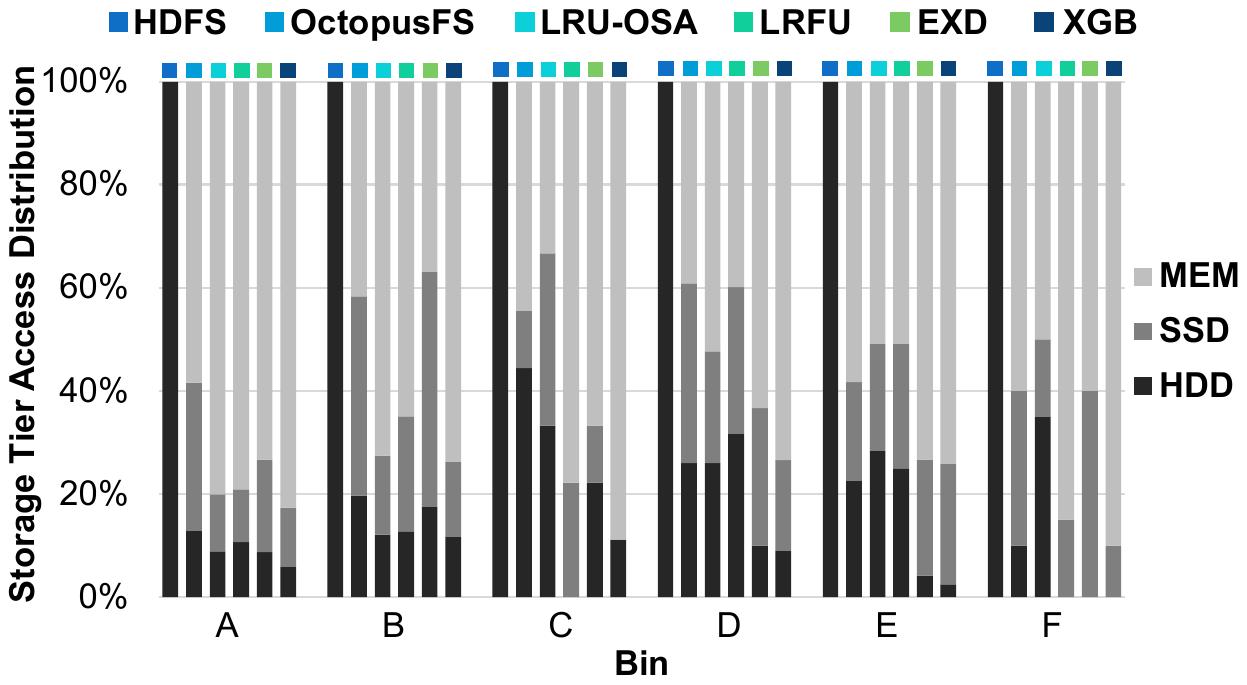}}
	\vspace{0.5ex}
	\caption{Access distributions for the storage tiers for the (a) FB and (b) CMU workloads}
	\label{fig:access-distribution}
\end{figure*}

\begin{figure}
	\centering
	\includegraphics[width=0.47\textwidth]{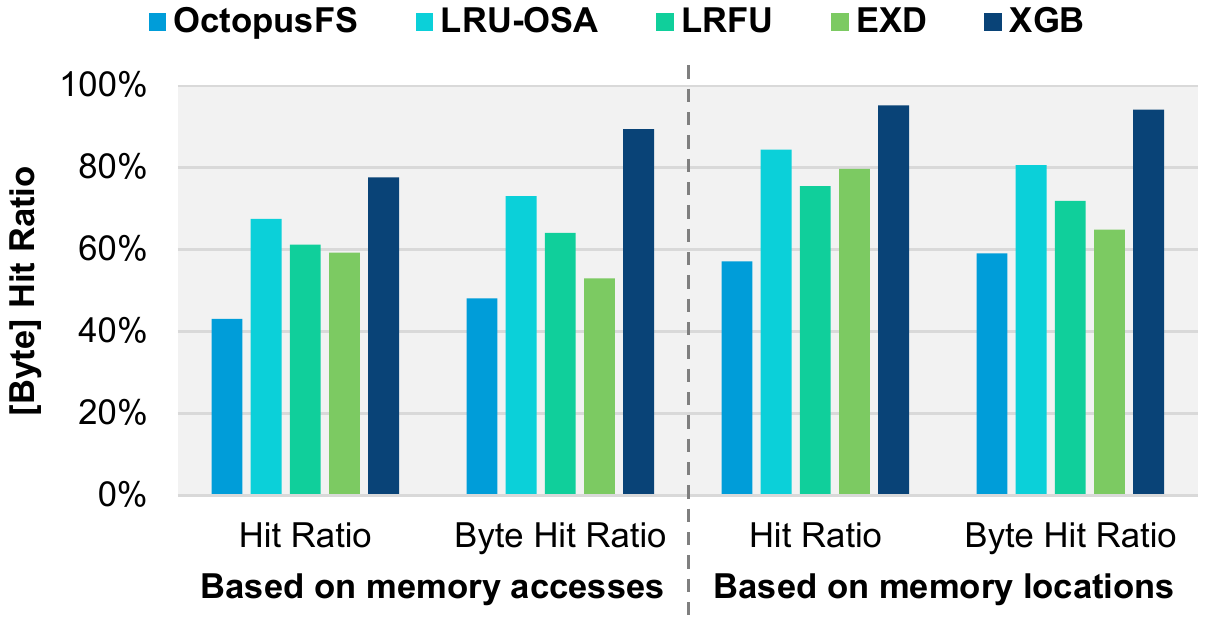}
	\vspace{0.5ex}
	\caption{Hit Ratio and Byte Hit Ratio for the FB workload based on memory accesses and locations}
	\label{fig:hit-ratios-both-policies-fb}
	\vspace{-2.5ex}
\end{figure}

Figure \ref{fig:completion-time} shows the reduction percentage in job completion time compared to the HDFS setting for both workloads for each bin. 
Small jobs (Bins A, B) experience only a small improvement in completion time for all policies, with less than 5\% and 10\% gain in FB and CMU, respectively.
This result is not surprising since the time spent in I/O 
is only a small fraction compared to CPU processing and scheduling overheads.
As the job input sizes increase, so do the gains in job completion time, while we start observing \textit{different behavior across the policies and between the two workloads}.
Specifically, the LRU-OSA and LRFU policies are performing quite well for the FB workload --- as it exhibits good temporal locality of reference --- resulting in up to 17\% reduction in completion time for large jobs (Bin F).
The CMU workload, on the other hand, has different access patterns that make the LRU-OSA and LRFU policies perform poorly, limiting gains down to 4-10\%.
In fact, LRU-OSA performs even worse than OctopusFS in some cases, which lacks any policy for automatically moving data across the storage tiers after the initial placement.

EXD performs well only for jobs in Bin D with 10\% gains in completion time, while it performs poorly for larger jobs in FB.
Recall that EXD explores the tradeoff between recency and frequency without taking file size into account.

In several cases, EXD downgrades large files before they are accessed, causing their data to be (re-)read from SSDs or HDDs.
Interestingly, EXD performs well for CMU, resulting in 13--16\% gains for Bins D and E.
Finally, our XGB policy is able to provide the \textit{highest reduction in average completion time across all job bins and for both workloads}.
For FB, there is a clear increasing gain as the job size gets larger with 18--27\% benefits, almost \textit{double} compared to the second-best policy.
For CMU, the gains are higher for medium jobs (Bins D, E) with benefits over 21\%, while the benefits for large jobs (Bin F) are still high at 15\%.
Overall, XGB is able to effectively learn the different access patterns and detect data reuse across jobs for both workloads.

Every time data is accessed from memory or even SSDs, the efficiency of the cluster improves.
Figure \ref{fig:cluster-efficiency} shows how this improvement in efficiency is derived from the different job bins.
Larger jobs have a higher contribution in efficiency improvement compared to small jobs since they are responsible for performing a larger amount of I/O (recall Table \ref{table:bins}).
Across the different policies, the trends for the efficiency improvement are similar to the trends for the completion time reduction discussed above:
LRU-OSA and LRFU generally offer good benefits for FB; EXD offers good benefits for CMU; and \textit{XGB offers the best gains in both workloads}.
Hence, improvements in cluster efficiency are often accompanied by lower job completion times, doubling the benefits.
In FB, for example, XGB is able to reduce the completion time of large jobs by 27\% while consuming 41\% less resources.

One interesting observation is that the magnitude of the gain in efficiency is higher than the magnitude of the reduction in completion times.
This is explained in two ways.
First, the jobs are executed as a set of parallel tasks.
Even if a large fraction of the tasks consume less resources via avoiding disk I/O, 
the remaining tasks may delay the completion of a job
(even though the benefits do propagate to all tasks due to lower I/O congestion).
Second, the job completion time also accounts for CPU processing as well as the output data generation, both of which are independent of the input I/O.
Nonetheless, improving cluster efficiency leaves significantly more room for executing more jobs and for better overlapping the background I/O generated by our policies.

To further understand the effects of the downgrade and upgrade policies, we look at the locations of the file block accesses, shown in Figure \ref{fig:access-distribution}.
Even though small jobs (Bins A, b) do not experience much benefit in either average completion time or cluster efficiency, we see a very high percentage of accesses (71-82\%) performed from memory, showing that all policies are able to retain reused small files in memory.
With larger files we observe a clear correlation between the number of accesses from memory and the performance gains; the higher the number of memory accesses, the higher the benefits for both efficiency and completion times.
Once again, \textit{XGB results in the highest percentage of memory accesses across all bins and workloads}, while also resulting in high access numbers from the SSD tier; confirming the efficacy of its downgrade and upgrade decisions.

Based on the number of tasks that accessed their input from memory, we compute two additional well-known metrics:
(i) \textit{Hit Ratio (HR)}, defined as the percentage of requests that can be satisfied by the memory tier; and
(ii) \textit{Byte Hit Ratio (BHR)}, defined as the percentage of bytes satisfied by the memory tier \cite{web-cache-survey-ijasca11}.
The results are shown in Figure \ref{fig:hit-ratios-both-policies-fb}.
We focus on the memory tier as it provides the highest benefits.
OctopusFS achieves less than 50\% for both ratios as less than half of the files have replicas in memory.
With a HR of \mytilde68\%, LRO-OSA and RFM performed about 8\% better compared to LRFU and EXD.
Finally, \textit{XGB offers the highest HR at 78\%}.
It is interesting to note that there is no clear correlation between HR and the performance results we saw in Section \ref{sec:eval:endtoend}, confirming that HR is not a good metric in the big data analytics settings \cite{pacman-nsdi12, big-sql-caching-socc16}.
The BHR, on the other hand, can help explain some of the results.
EXD has a low BHR of 53\%, which indicates that a significant fraction of tasks did not read their data from memory, leading to the low cluster efficiency observed in Figure \ref{fig:cluster-efficiency}.
On the other hand, LRU-OSA and RFM scored high on both BHR and cluster efficiency for most job bins.
\textit{XGB is able to achieve the highest BHR at 94\%}, showcasing the ability of XGB to keep the most relevant files in memory.

Even though a file replica exists in memory, it is not guaranteed that it will be accessed from memory because the current schedulers of big data platforms (e.g., Hadoop, Spark) do not account for the presence of multiple storage tiers in the cluster.
Hence, we also looked at the storage location of each file right before it was accessed and computed the corresponding HR and BHR, also shown in Figure \ref{fig:hit-ratios-both-policies-fb}.
Overall, we noticed that HR and BHR based on location were 15-20\% and 5-10\% higher compared to HR and BHR based on accesses, respectively.
The results motivate further research on how higher-level systems should exploit the presence of multiple storage tiers to further improve workload performance and cluster utilization.

\begin{figure}[t!]
	\centering
	\includegraphics[width=0.47\textwidth]{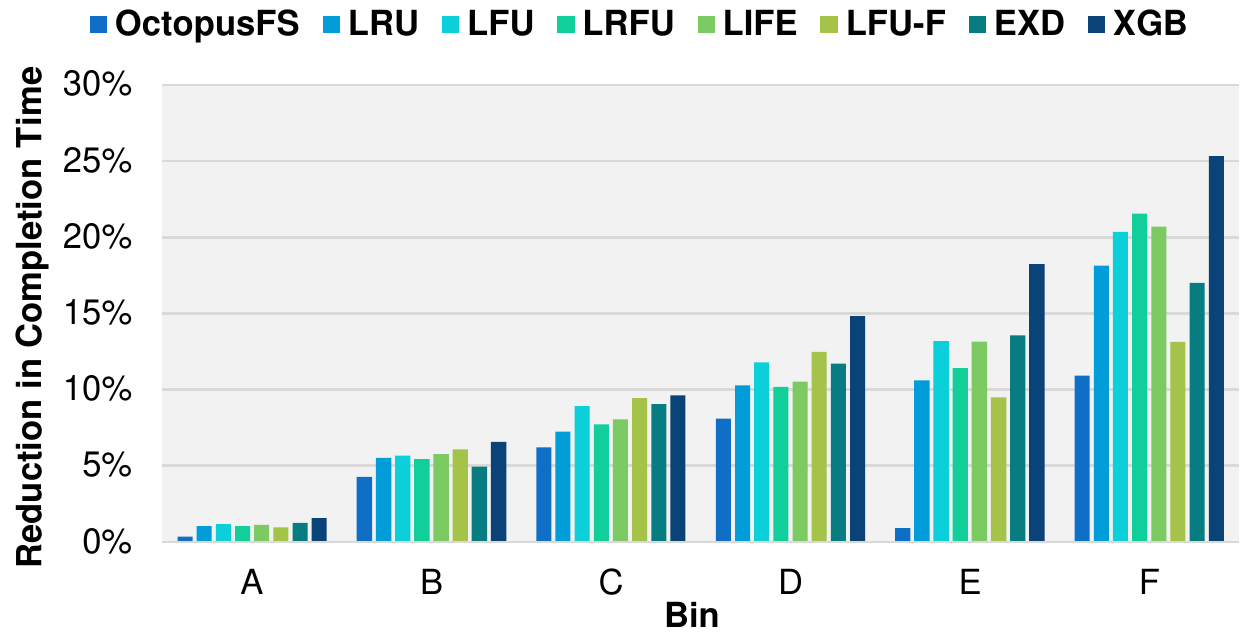}
	\vspace{1ex}
	\caption{Percent reduction in completion time for downgrade policies over HDFS for the FB workload}
	\label{fig:downgrade-results-fb}
	\vspace{-1ex}
\end{figure}

\subsection{Comparison of Downgrade Policies}
\label{sec:eval:downgrade}

\begin{figure}[t!]
	\centering
	\includegraphics[width=0.4\textwidth]{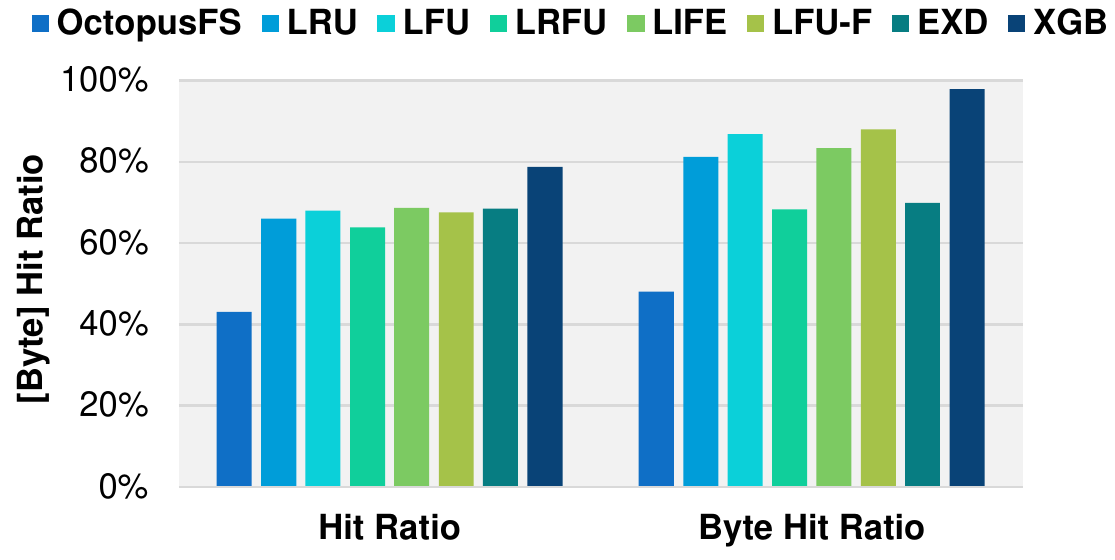}
	\vspace{1ex}
	\caption{Hit Ratio and Byte Hit Ratio for downgrade policies for FB based on memory accesses}
	\label{fig:downgrade-hit-ratios-fb}
	\vspace{-2ex}
\end{figure}

For this experiment, we executed the two workloads over Octopus++ using all downgrade policies listed in Table \ref{table:downgrade-policies}, while disabling upgrades.
Our goals are (i) to isolate the effects of downgrading from upgrading, and (ii) to gain additional insights from the downgrade policies.
Figure \ref{fig:downgrade-results-fb} shows the percent reduction in average completion time for all jobs in FB broken down into bins.
The trends for the policies we already studied in Section \ref{sec:eval:endtoend} are similar in general.
However, a careful comparison between Figures \ref{fig:completion-time}(a) and \ref{fig:downgrade-results-fb} yields some interesting observations.
First, the LRU and LRFU downgrade policies cause the same amount of gains in completion time as when upgrade policies are present.
The EXD upgrade policy has a strong negative effect on the completion time benefits compared to only enabling the downgrade policy (5--14\% lower gains for large jobs).
The extra amount of data upgraded is causing the downgrade of other useful files, negatively affecting performance.
The pairing of the XGB downgrade and upgrade policies has a small positive affect of 1.5-3.6\% additional reduction in completion times compared to using the XGB downgrade policy alone.
This highlights the effectiveness of both XGB policies in making the right decisions when moving files across the storage tiers.

The LIFE policy \cite{pacman-nsdi12}, which is designed to improve job completion time, performs fairly well in this metric, especially for larger jobs (13--21\% gains for Bins E, F).
However, it falls short compared to the 18--25\% benefits provided by XGB for Bins E and F.
LRF-U \cite{pacman-nsdi12}, which targets at improving cluster efficiency, ranks second in this metric for small to medium jobs (Bins B-D) but performs poorly for larger jobs.
Recall that LRF-U does not take file size into account.
\textit{XGB, on the other hand, provides the highest gains in cluster efficiency for all bins}.
The CMU results do not provide any additional insights and are not presented due to space constraints.

To further analyze the performance of the downgrade policies we also computed the  Hit Ratio (HR) and Byte Hit Ratio (BHR).
We focus on the memory tier as it provides the highest benefits.
Figure \ref{fig:downgrade-hit-ratios-fb} shows the HR and BHR of all downgrade policies based on memory accesses.
OctopusFS achieves less than 50\% for both ratios as less than half of the files have replicas in memory.
With the exception of XGB, all other policies achieve a similar HR of \mytilde67\%.
BHR reveals a different picture with LRFU and EXD offering \mytilde69\% BHR (which explains their lower gains in cluster efficiency), while the rest offer \mytilde85\%.
\textit{XGB is able to achieve a 98\% BHR, highlighting the policy's ability to maintain the most relevant files in memory.}

\subsection{Comparison of Upgrade Policies}
\label{sec:eval:upgrade}

\begin{figure}[t!]
	\centering
	\includegraphics[width=0.47\textwidth]{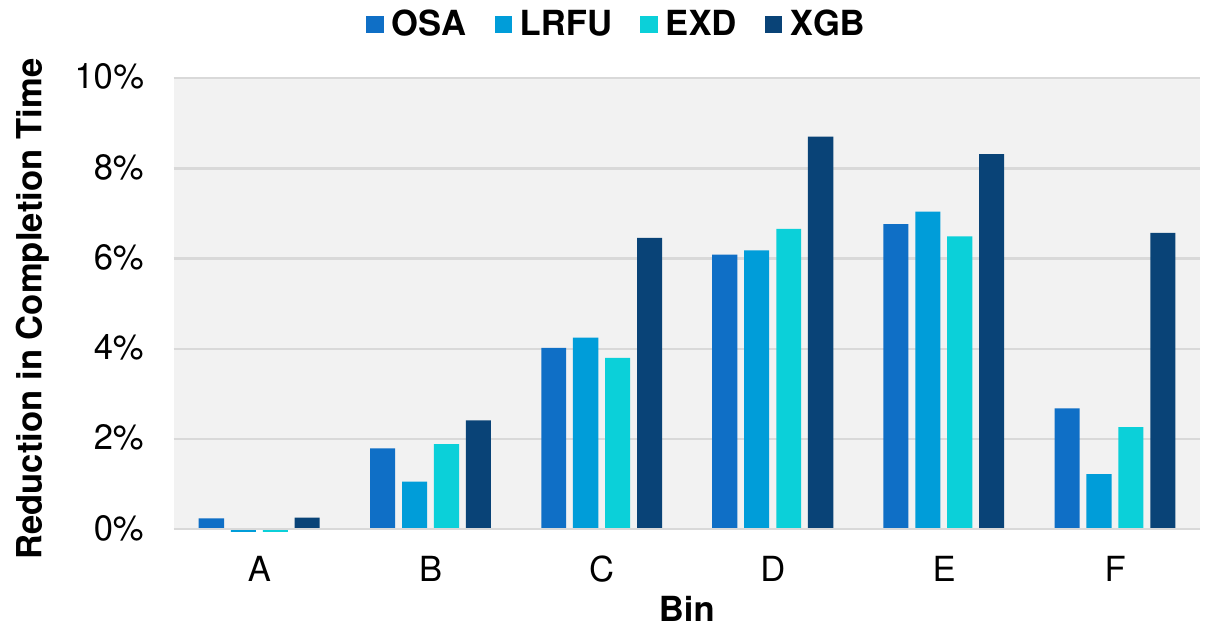}
	\vspace{1ex}
	\caption{Percent reduction in completion time for upgrade policies over HDFS for the FB workload}
	\label{fig:upgrade-results-fb}
	\vspace{-1ex}
\end{figure}

\begin{table}[t]
	\setlength{\tabcolsep}{2.5pt}
	\centering
	\caption{Statistics for upgrade policies for FB}
	\vspace{1ex}
	\renewcommand{\arraystretch}{1.1}
	\small
	\begin{tabular}{| l | c | c | c | c |}
		\hline
		Policies	&	GB Read from	&	GB Upgraded to	&	Byte	&	Byte \\
		&	Memory Tier	&	Memory Tier	&	Accuracy	&	Coverage	\\
		\hline
		OSA	&	9.41	&	34.52	&	0.27	&	0.21	\\
		LRFU	&	9.03	&	22.82	&	0.40	&	0.21	\\
		EXD	&	6.45	&	22.59	&	0.29	&	0.15	\\
		XGB	&	13.77	&	27.66	&	0.50	&	0.31	\\
		\hline
	\end{tabular}
	\label{table:upgrade-only-stats-fb}
	\vspace{-1ex}
\end{table}

\begin{figure*}[t!]
	\centering
	\includegraphics[width=0.47\textwidth]{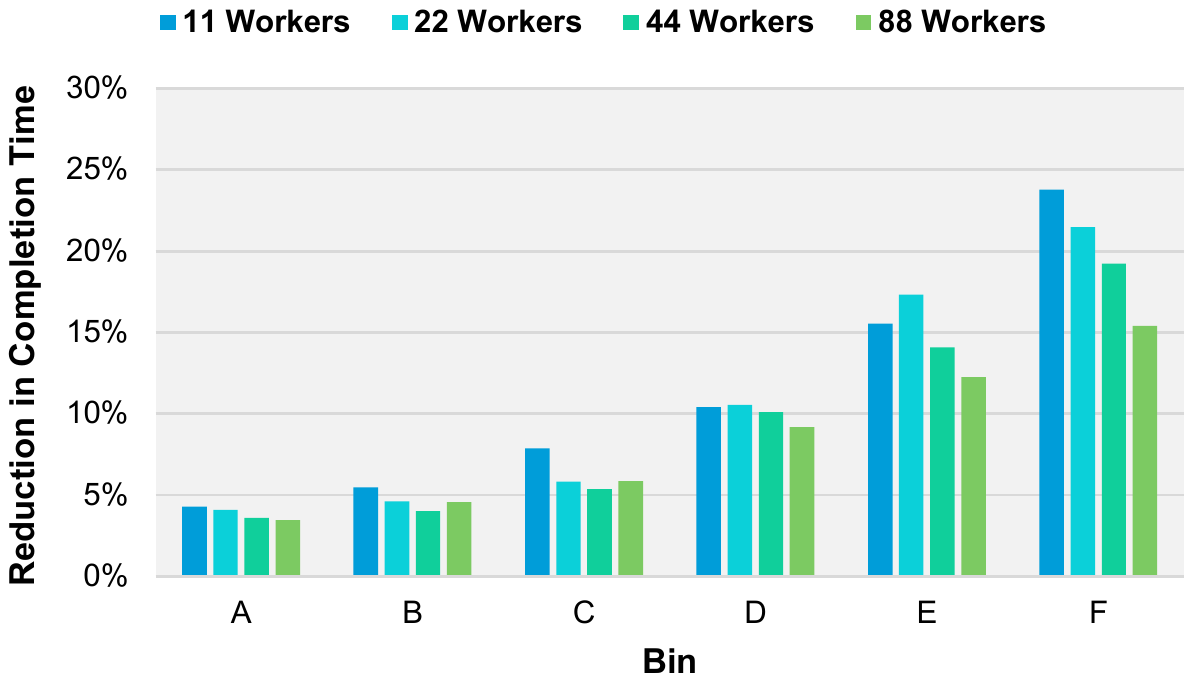}
	\hfil
	\includegraphics[width=0.47\textwidth]{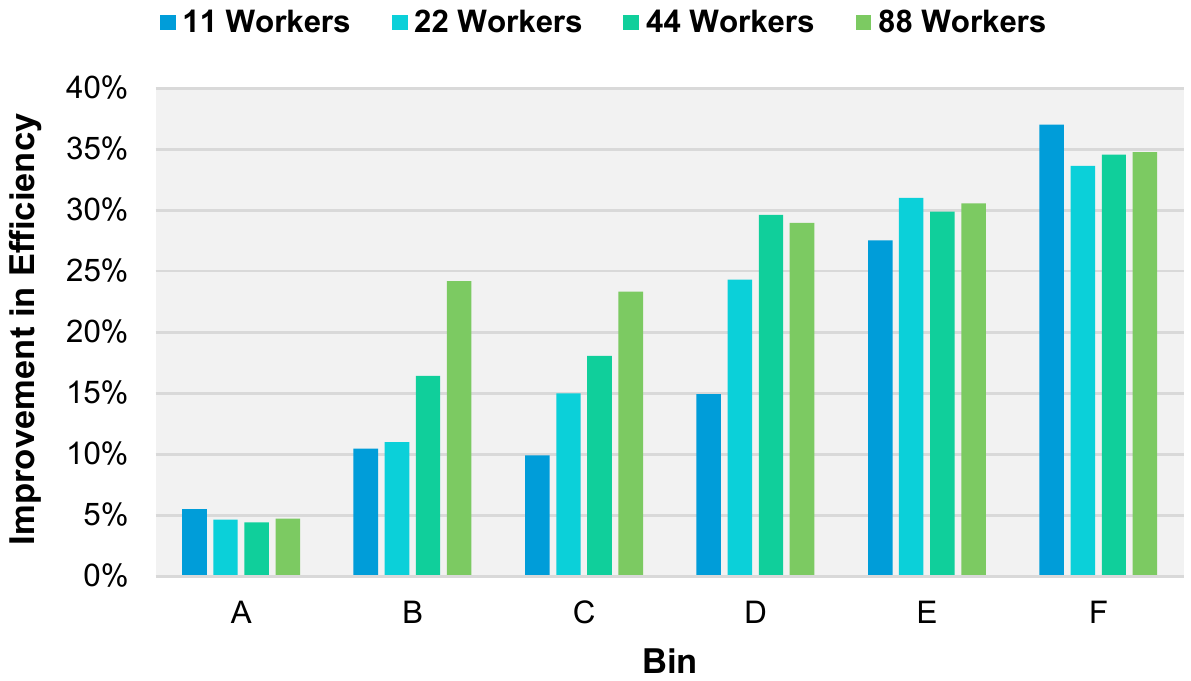}
	\vspace{1ex}
	\caption{(a) Percent reduction in completion time and (b) percent improvement in cluster efficiency over HDFS for the FB workload when using the XGB downgrade/upgrade policies as we scale the cluster size on Amazon EC2}
	\label{fig:ec2-results}
	\vspace{-1.5ex}
\end{figure*}

In this section, we evaluate all upgrade policies listed in Table \ref{table:upgrade-policies} in isolation.
We instructed the data placement policy of Octopus++ to initially place all file replicas on the HDD tier and let the upgrade policies decide when to move replicas to higher tiers.
Figure \ref{fig:upgrade-results-fb} shows the gains in job completion time for each upgrade policy for each job bin.
Overall, the gains are limited to less than 9\% since gains are only possible when an input file is accessed repeatedly, provided that a policy was able to upgrade the file early on.
OSA, the simple policy that upgrades a file into memory upon access, performs relatively well for most bins (with 2--7\% gains), while the other policies offer similar benefits in most cases.
The only policy that stands apart is the XGB one, which offers the highest benefits, showcasing the predictive powers of our ML model.

Table \ref{table:upgrade-only-stats-fb} lists two insightful statistics --- amount of data upgraded to and read from memory --- as well as two important web prefetching metrics:
(1) \textit{Byte Accuracy (BAc)}, defined as the ratio of data read from memory to the total amount of data upgraded; and
(2) \textit{Byte Coverage (BCo)}, defined as the ratio of data read from memory to the total amount of data read \cite{web-cache-survey-ijasca11}.
As the least selective policy, OSA upgraded the largest amount of data (34.5GB) leading to the $2^{nd}$ highest amount of data read from memory (9.4GB) but with a fairly low BAc.
LRFU and EXD upgraded about the same amount of data (\mytilde22.7GB) but only LRFU was able to score a relatively high BAc of 40\%.
Even though XGB upgraded slightly more data (27.6GB), it was able to achieve a higher BAc at 50\% and the highest BCo at 31\%, explaining the high performance benefits attained by XGB.

\subsection{Scalability Experiments}
\label{sec:eval:scalability}

We repeated our experiments on Amazon EC2 using the m4.2xlarge instance type (8 cores, 32GiB RAM) with one SSD and three HDD attached EBS volumes per Worker node to resemble our local cluster setup.
We scaled up the EC2 cluster from 11 to 88 worker nodes (plus 1 master node) while proportionally increasing the workload data sizes and we were able to fully replicate our results.
In particular, the gains in both completion time and cluster efficiency while using the XGB downgrade and upgrade policies increase as the job sizes in the FB workload increase and have similar values as the ones observed on our local cluster (recall Section \ref{sec:eval:endtoend}).

Figures \ref{fig:ec2-results}(a) and \ref{fig:ec2-results}(b) show the percent reduction in completion time and percent improvement in cluster efficiency over HDFS, respectively, as we scale the cluster size on Amazon EC2.
Based on these results, we extract two key insights.
First, the improvement in cluster efficiency increases with the cluster size, especially for small-medium jobs (Bins B-D).
For example, the improvement in efficiency increases from 10\% to 23\% for jobs in Bin C as we increase the number of Worker nodes from 11 to 88, revealing the increasing benefits of avoiding disk I/O and better utilizing the cluster resources.
Second, the gains in completion time are similar for small-medium jobs but decrease for large jobs (from 24\% to 15\% gains in Bin F) as the cluster size increases.
As we scale up the cluster size, we also scale the workload data sizes, including the jobs' output data size.
Hence, even though the benefits from saving input data I/O increase with cluster size, the output data I/O increases dis-proportionally as the output files have a replication factor of 3, leading to the observed lower benefits for large cluster sizes.

\begin{figure}[t!]
	\centering
	\includegraphics[width=0.47\textwidth]{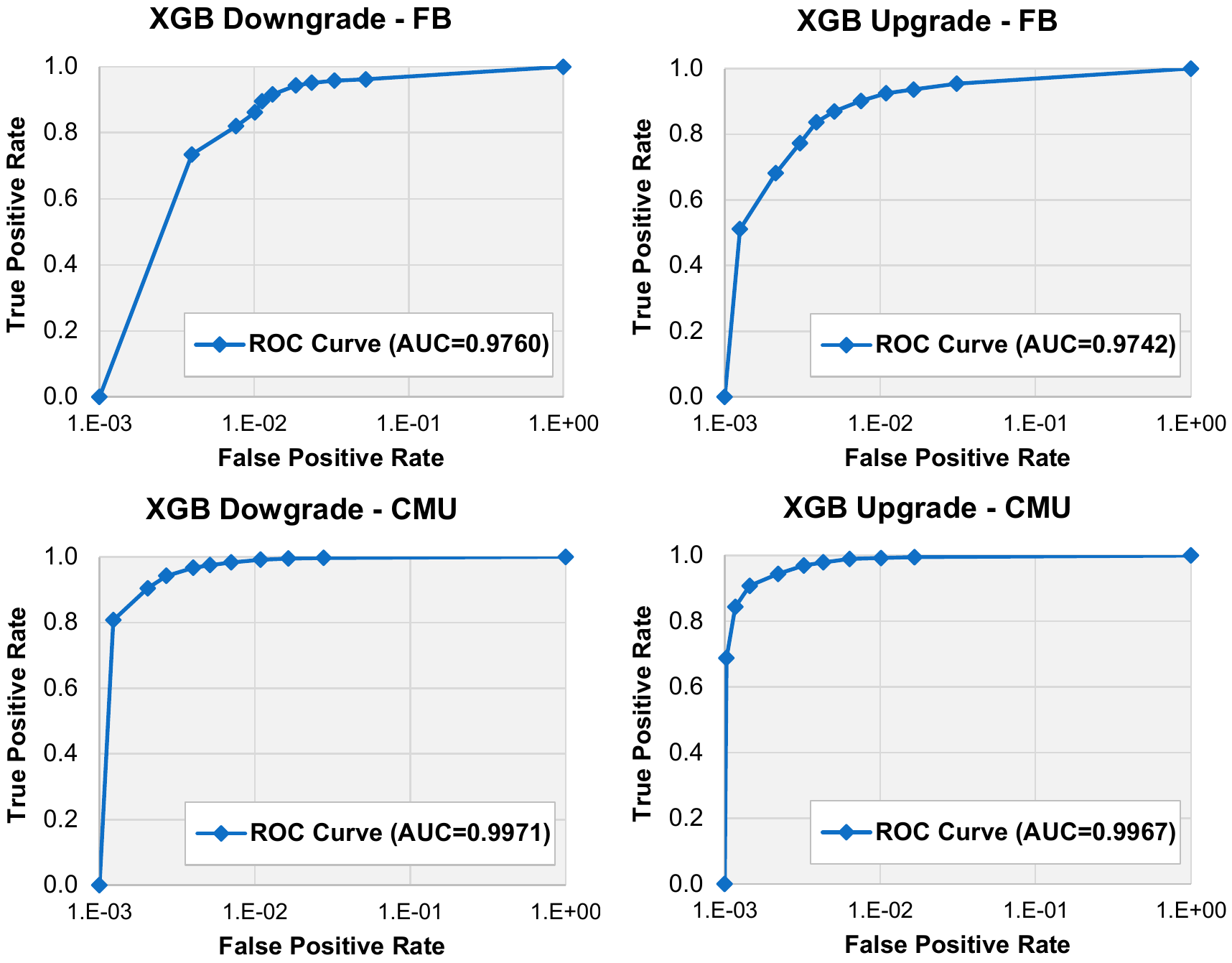}
	\vspace{1ex}
	\caption{ROC curves for XGB downgrade/upgrade}
	\label{fig:roc-curves}
	\vspace{-1ex}
\end{figure}

\subsection{XGBoost Model Evaluation}
\label{sec:eval:xgboost}

\begin{figure*}[t]
	\begin{minipage}[t]{0.322\linewidth}
		\vspace{0pt}
		\centering
		\includegraphics[width=.98\textwidth]{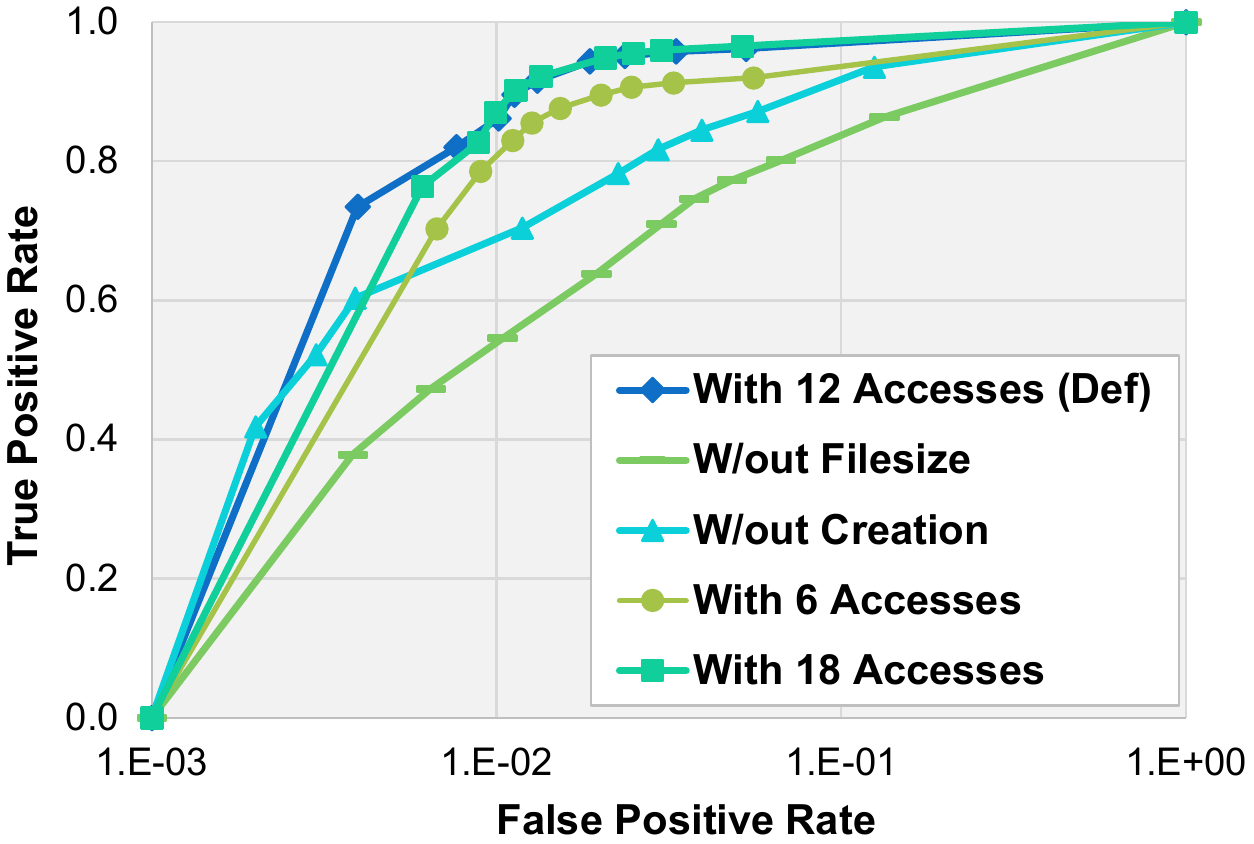}
		\vspace{3pt}
		\caption{ROC curves for FB downgrade XGBoost models with selected features}
		\label{fig:roc-curves-features}
	\end{minipage}
	\hfill
	\begin{minipage}[t]{0.319\textwidth}
		\vspace{0pt}
		\centering
		\includegraphics[width=0.98\textwidth]{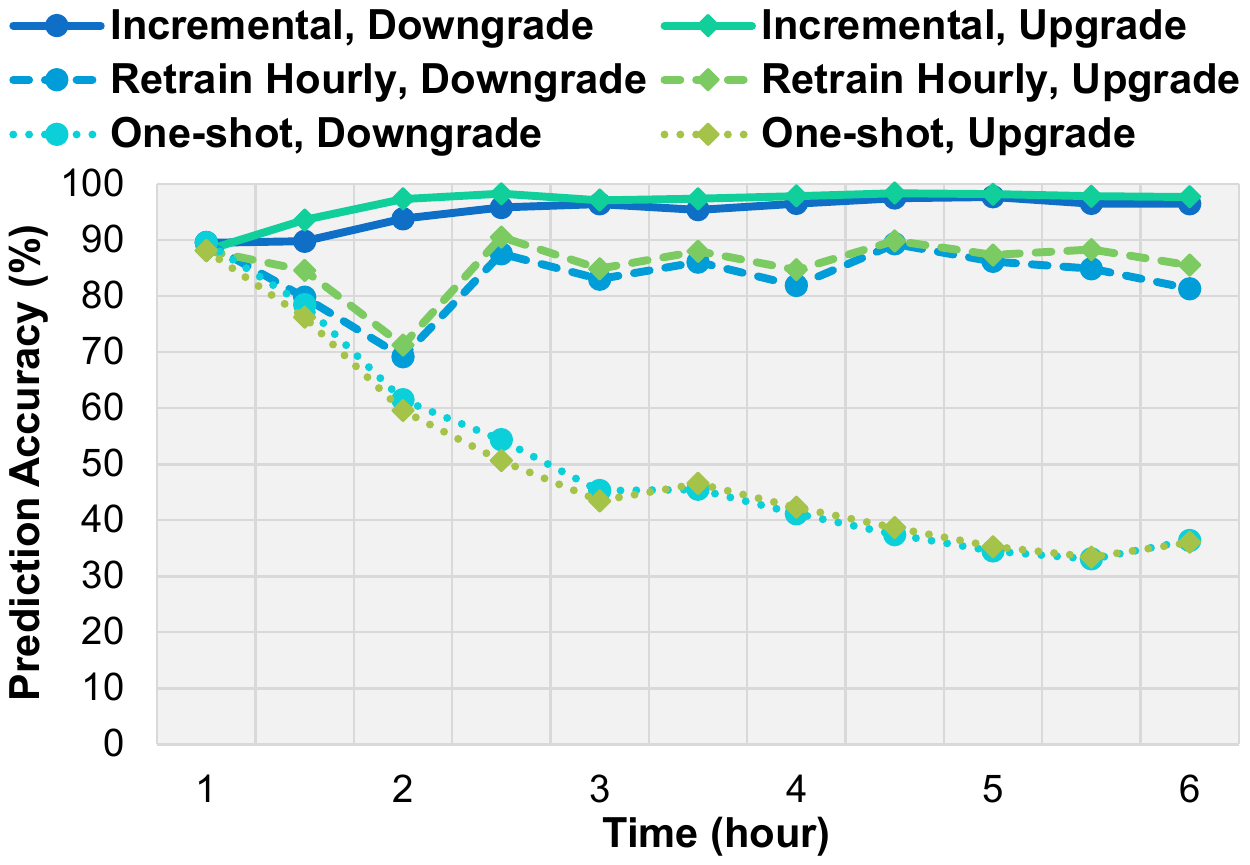}
		\vspace{1pt}
		\caption{Prediction accuracies of the incremental learning, retrain hourly, and one-shot approaches for FB}
		\label{fig:incremental-learning}
	\end{minipage}
	\hfill
	\begin{minipage}[t]{0.319\linewidth}
		\vspace{0pt}
		\centering
		\includegraphics[width=.98\textwidth]{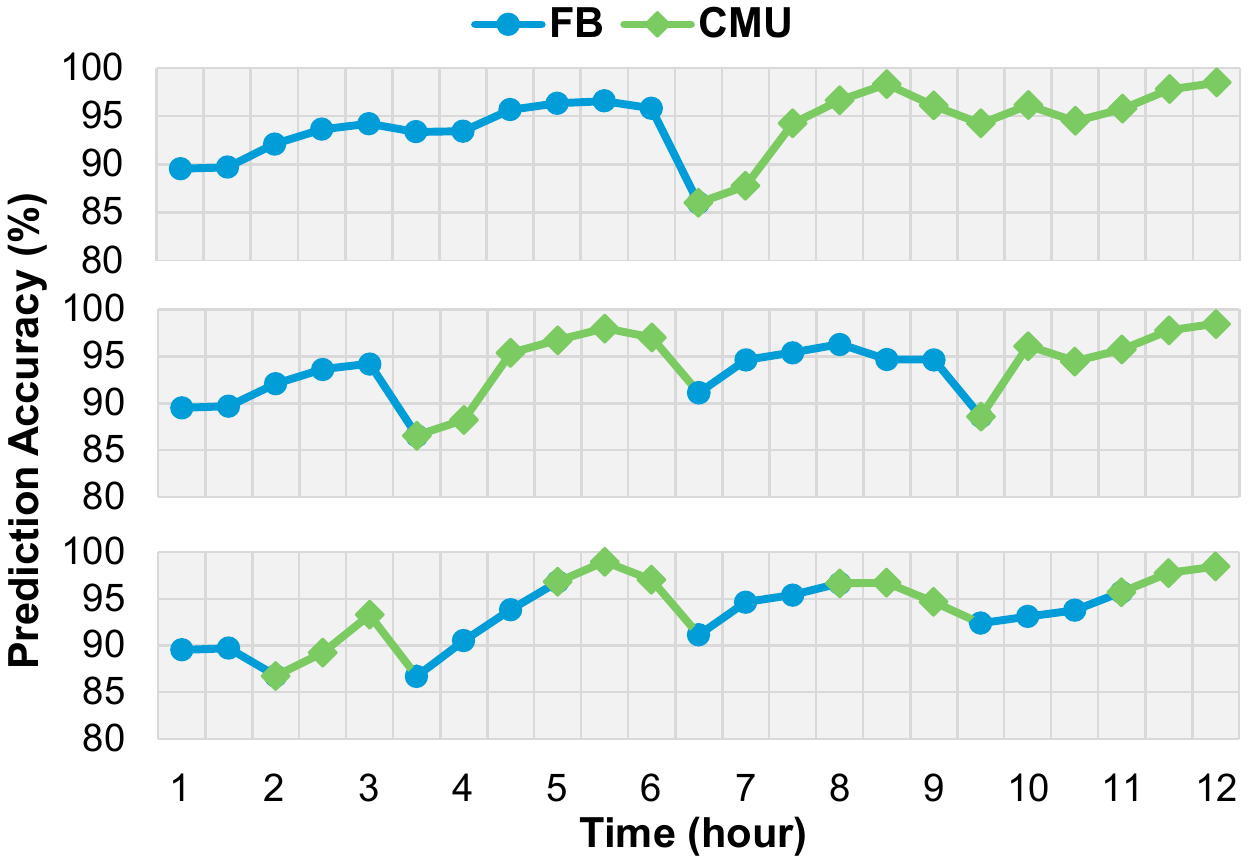}
		\vspace{1pt}
		\caption{Prediction accuracies of downgrade XGBoost models when alternating FB and CMU workloads}
		\label{fig:mix-learning}
	\end{minipage}
	\vspace{-2ex}
\end{figure*}

We evaluate the performance of our XGBoost models using a receiver operating characteristic (ROC) curve and the area under the curve (AUC) \cite{info-retrieval-book}.
The ROC curve takes as input the file access probabilities predicted by the model and the true class labels.
It then plots the true positive rate (i.e., the probability of detection) against the false positive rate (i.e., the probability of false alarm) at various threshold settings.
To train our models and perform a proper out-of-sample analysis, we split our 6-hour data set into training (first 4 hours), validation (5th hour), and test (6th hour) sets.
All results presented in this Section are based on test data.
The FB and CMU data sets consist of 17968 and 13920 data points, respectively, evenly distributed (approximately) in each hour.
Figure \ref{fig:roc-curves} shows the 4 ROC curves for the downgrade and upgrade models trained and evaluated over the FB and CMU workloads (note the logarithmic scale of the x-axis).
In all cases, the curves are near point $(0,1)$ with AUC values higher than 0.97 (1 is the max), which indicate the very high prediction performance of our models.
The prediction accuracy at the chosen discrimination threshold (0.5) is between 97-99\% in all four cases.

The features used by our XGBoost model formulation are the file size, the creation time, and the last $k$ (default=12) access times.
To assess the impact of the selected features, we evaluated the performance of the XGBoost model while varying the features used.
In particular, Figure \ref{fig:roc-curves-features} shows the ROC curves for the FB downgrade case (other cases are similar) when the features
(i) do not include the file size;
(ii) do not include the creation time;
(iii) use only the last 6 access times;
(iv) use the default 12 access times; and
(v) use 18 access times.
Note that (iii)-(v) do use the file size and creation time.
The results reveal that both the file size and creation time are individually important predictors of file accesses and improve the overall performance of the model.
Using only 6 time accesses still leads to good model performance, albeit slightly lower than our default case, while using 18 time accesses has a marginal impact.
Overall, these results verify the selection of the aforementioned features to be used for predicting file access patterns.

Next, we evaluate our \textit{incremental learning} approach that cumulatively trains a model over time against 
(i) retraining the model every one hour and
(ii) a \textit{one-shot} learner that trains on data once from the first hour.
Figure \ref{fig:incremental-learning} shows how the prediction accuracy (i.e., the ratio of true predictions over all of them) for the three approaches varies over time for the downgrade and upgrade policies over FB.
Although a one-shot learner may start with a high initial accuracy near 90\%, over time it leads to a significant degradation of accuracy below 40\% as the workload evolves.
The retraining approach exhibits an oscillating patter of accuracy that increases right after each training but stays within 80\% and 90\%.
On the contrary, the incremental learner becomes better over time, efficiently adapting to new workloads,
and continues to produce accurate predictions (\mytilde98\%) over the entire duration of our experiments.

Finally, we investigate the impact of sudden access pattern changes to the XGBoost incremental model performance.
Figure \ref{fig:mix-learning} shows the prediction accuracy of the XGB downgrade policy over time as we switch between the FB and CMU workloads.
In the first experimental variation, we execute FB for 6 hours and then switch to CMU, which exhibits different access patterns.
At that point, accuracy drops 9.8\% down to 86\% but then quickly increases to over 95\% as the model starts learning the new access patterns.
In the next 2 variations, we alternate the FB and CMU execution every 3 and 1.5 hours, respectively, and observe that
(i) as time goes by, the drops in accuracy decrease in magnitude,
(ii) the more we interleave the workloads, the lower the drop as the model learns both workloads, and
(iii) the model is always able to learn and increase accuracy quickly.

\subsection{System Overheads}
\label{sec:eval:overheads}

Adding one training sample in an XGBoost model takes on average 0.16ms, while making a prediction takes 1.8ns.
Overall, during a 6-hour-long experiment, model training accounted for 5.3 CPU seconds in total, while selecting a file to downgrade or upgrade amounted to 0.49 CPU seconds; showcasing the negligible CPU overhead caused by XGBoost.
In terms of memory, an XGBoost model consumes \mytilde200KB.
In addition, we maintain a max of 956 bytes per file (at most 12 access times plus some auxiliary data) for generating the feature vectors, which in our experiments added just a few MB of memory.
Even in large clusters with millions of files, the extra memory overhead is $<$1GB, which is a fraction of the total memory needed by an HDFS NameNode at that scale and justifiable given the attainable performance benefits.

\section{Conclusions}
\label{sec:conclusions}

This paper describes a framework for automatically managing data across storage tiers in distributed file systems (DFSs) using a set of pluggable policies.
The generality of the framework is evident by the 11 downgrade and upgrade policies implemented based on both old and new techniques.
Our proposed policies employ light-weight gradient boosted trees for learning how files are accessed by a workload and use that information to make decisions on which files to move up or down the tiers.
The models are incrementally updated based on how the file access patterns change and so are able to maintain high prediction accuracy over time.
The framework and all policies have been implemented in a real distributed file system and successfully evaluated over two realistic workloads.
Our evaluation results also motivate further research on how higher-level systems should exploit the presence of multiple storage tiers to further improve performance and cluster utilization.

\balance

\section{Acknowledgments}
This work was supported by the AWS Cloud Credits for Research program.

\bibliographystyle{abbrv}
\bibliography{paper} 

\end{document}